\documentclass[12pt,preprint]{aastex}

\usepackage{emulateapj5} 
\newcommand{\et}{et al.}

\newcommand{\nh}{N_{\rm H}} 
\newcommand{\Msun}{\hbox{$\rm\thinspace M_{\odot}$}}
\slugcomment{}
\shorttitle{Suzaku Cen A observation}
\shortauthors{Markowitz et al.}
\begin{document}
\title{The Suzaku Observation of the Nucleus of the 
Radio-Loud Active Galaxy Centaurus A: Constraints 
on Abundances of the Accreting Material}

\author{A. Markowitz\altaffilmark{1,2},       
T. Takahashi\altaffilmark{3},
S. Watanabe\altaffilmark{3},
K. Nakazawa\altaffilmark{3},
Y. Fukazawa\altaffilmark{4},
M. Kokubun\altaffilmark{5},       
K. Makishima\altaffilmark{5,6},   
H. Awaki\altaffilmark{7},         
A. Bamba\altaffilmark{6},         
N. Isobe\altaffilmark{6},         
J. Kataoka\altaffilmark{8},       
G. Madejski\altaffilmark{9,10},     
R. Mushotzky\altaffilmark{1},    
T. Okajima\altaffilmark{1}, 
A. Ptak\altaffilmark{11}           
J.N. Reeves\altaffilmark{1,11},    
Y. Ueda\altaffilmark{12},  
T. Yamasaki\altaffilmark{4},
T. Yaqoob\altaffilmark{11}  
\altaffiltext{1}{X-ray Astrophysics Laboratory, Code 662, NASA/Goddard Space Flight Center, Greenbelt, MD 20771, USA; agm@milkyway.gsfc.nasa.gov}
\altaffiltext{2}{NASA Post-doc Research Associate}
\altaffiltext{3}{Institute of Space and Astronautical Science, JAXA, Sagamihara, Kanagawa 229-8510, Japan}
\altaffiltext{4}{Department of Physics, Hiroshima University, 1-3-1 Kagamiyama, Higashi-Hiroshima, 739-8526, Japan}
\altaffiltext{5}{Department of Physics, Graduate School of Science, University of Tokyo, Hongo 7-3-1, Bunkyo, 113-0033, Japan} 
\altaffiltext{6}{The Institute of Physical and Chemical Research (RIKEN), 2-1 Hirosawa, Wako, 351-0198, Japan}
\altaffiltext{7}{Department of Physics, Faculty of Science, Ehime University, Bunkyo-cho, Matsuyama, Ehime 790-8577, Japan}
\altaffiltext{8}{Department of Physics, Tokyo Institute of Technology, Meguro, Tokyo 152-8551, Japan}
\altaffiltext{9}{Stanford Linear Accelerator Center (SLAC), 2575 Sand Hill Road, Menlo Park, CA 94025, USA}
\altaffiltext{10}{Kavli Institute for Particle Astrophysics and Cosmology, Stanford University, Stanford, CA 94305, USA}
\altaffiltext{11}{Department of Physics and Astronomy, Johns Hopkins University, Baltimore, MD 21218, USA}
\altaffiltext{12}{Department of Astronomy, Kyoto University, Sakyo-ku, Kyoto 606-8502, Japan}
}


\begin{abstract}
A {\it Suzaku} observation of the nucleus of the radio-loud AGN Centaurus A 
in 2005 has yielded a broadband spectrum spanning 0.3 to 250 keV. The net 
exposure times after screening were: 70 ks per X-ray Imaging Spectrometer 
(XIS) camera, 60.8 ks for the Hard X-ray Detector (HXD) PIN, and 17.1 ks for 
the HXD GSO. The hard X-rays are fit by two power laws of the same slope, 
absorbed by columns of 1.5 and 7 $\times$ 10$^{23}$ cm$^{-2}$ respectively.
The spectrum is consistent with previous suggestions that the power-law 
components are X-ray emission from the sub-pc VLBI jet and from Bondi 
accretion at the core, but it is also consistent with a partial-covering 
interpretation. The soft band is dominated by thermal emission from the 
diffuse plasma and is fit well by a two-temperature {\sc vapec} model,
plus a third power-law component to account for scattered nuclear emission, 
jet emission, and emission from X-ray Binaries and other point sources. 
Narrow fluorescent emission lines from Fe, Si, S, Ar, Ca and Ni are detected.
The Fe K$\alpha$ line width yields a 200 light-day lower limit on the 
distance from the black hole to the line-emitting gas. Fe, Ca, and S K-shell 
absorption edges are detected. Elemental abundances are constrained via 
absorption edge depths and strengths of the fluorescent and diffuse plasma 
emission lines. The high metallicity ([Fe/H]=+0.1) of the circumnuclear material
suggests that it could not have originated in the relatively metal-poor outer 
halo unless enrichment by local star formation has occurred. Relative abundances 
are consistent with enrichment from Type II and Ia supernovae.
\end{abstract}
                                                                         

\section{Introduction}

The radio-loud active galactic nucleus (AGN) Centaurus A 
(NGC 5128) is one of the most extensively studied AGNs at all 
wave bands, thanks to its proximity (distance of $3.8 \pm 0.4$ Mpc, 
Rejkuba 2004; 1$\arcmin$ = 1 kpc) and brightness 
(2--10 keV flux typically $\sim$2 $\times$10$^{-10}$ erg cm$^{-2}$ s$^{-1}$). 
Cen A features the nearest AGN jet and is considered a prototypical FR I radio galaxy.
Its radio structure consists of extended kpc-scale
outer, middle, and inner radio lobes, fed by a 
one-sided kpc-scale jet. The jet is at an angle of $\sim$60$\degr$
to the line of sight, leading to its description as
a 'misaligned' BL Lac object (e.g., Bailey \et\ 1986).
VLBI observations have also revealed a sub-pc jet 
and unresolved ($\sim0.1$pc) core (Tingay \et\ 1998; see also
Israel 1998 for a review).

The host galaxy of Cen A 
is a giant elliptical with a pronounced optical dust lane
that obscures the inner few kpc and is believed to be an edge-on 
disk structure (Quillen \et\ 1992). 
The inner dust disk, with associated young stars, H II regions, and other
ionized gas, is the site of vigorous star formation, e.g., 
observed along the edges of the inner disk (e.g., Ebneter \& Balick 1983, Dufour \et\ 1979).
The dust disk is interpreted as the result of a merger 
between a large elliptical and a gas-rich disk galaxy at least a Gyr ago
(e.g., Israel 1998). 
Infrared studies are necessary to reveal the compact, pc-scale 
core (e.g., Karovska \et\ 2003). NIR studies (e.g, Schreier \et\ 1998) 
have revealed a circumnuclear disk tens--hundreds of pc wide, surrounding
the central black hole. Stellar kinematic
studies indicate a black hole mass near 2$\times$10$^8$ $\Msun$ (Silge \et\ 2005)

Previous studies of the nuclear X-ray emission from $\sim$4 keV to several hundred keV
(e.g., Rothschild \et\ 1999) have established the presence of a non-thermal,
power-law continuum whose origin is not certain. It could be a signature of accretion,
or it could be associated with jet emission processes, e.g., synchrotron 
or inverse Compton emission from the sub-pc VLBI jet. 

Below $\sim$4 keV, the X-ray continuum 
emission undergoes moderately heavily absorption, although the nature of
the X-ray obscuring material is not certain; i.e., it could have a sky-covering fraction
of 1 as seen from the central X-ray source or it could be in the form of a dusty torus. Rothschild
\et\ (2006) have noted $\sim$50$\%$ variations in the absorbing column over 
$\sim$20 years, and suggested that the nucleus is seen through the edges of a 
warped, rotating disk. The strong X-ray absorption makes Cen A relatively unique: the vast
majority of FR I radio galaxies possess relatively weak X-ray absorbing 
columns ($\lesssim$ 10$^{20-21}$ cm$^{-2}$). Heavy X-ray absorption, 
meanwhile, is usually seen only in FR II radio galaxies, although it is 
not yet clear whether the dichotomy in X-ray absorption properties has 
the same origin as the FR I/II dichotomy, e.g., differing accretion modes
(Donato \et\ 2004, Balmaverde \et\ 2006, Evans \et\ 2006).

Previous X-ray spectral fits to the nucleus of Cen A center on one or more
power laws, with varying degrees of absorption. For instance,
using {\it ROSAT} and {\it ASCA}, Turner \et\ (1997) suggested
that one plausible description of the 
nuclear emission is a partial-covering model, wherein
the nucleus is seen through three layers of absorption:
40$\%$, 59$\%$, and 1$\%$ of the nuclear emission is
obscured by columns of 4, 1 and 0.01 $\times$ 10$^{23}$ cm$^{-2}$, respectively.
Using {\it Chandra}-HETGS and {\it XMM-Newton} observations,
Evans \et\ (2004) suggested a different interpretation. They
fit the hard X-ray continuum spectrum with a heavily-absorbed
power law, thought to be associated with Bondi accretion at the core, plus
a relatively less-absorbed power law thought to be associated with the sub-pc VLBI jet,
although they did not rule out the partial-covering interpretation.

There is a prominent Fe K$\alpha$ emission line at 6.4 keV,
first noted by Mushotzky \et\ (1978). It is narrow (e.g., 
FWHM = 2200$\pm$900 km s$^{-1}$, Evans \et\ 2004), and its 
flux is historically nearly constant despite long-term continuum variations 
spanning a factor of $\sim$5 over 20 years (Rothschild \et\ 1999), 
implying an origin for the line that is distant from the origin of the variable continuum. 

The soft X-ray emission within $\sim$6 kpc of the nucleus of
Cen A consists of both diffuse and point-like emission.
There is diffuse emission associated with the kpc-scale jet 
extending from $<$ 60 pc to the NE radio lobe (e.g., Kraft \et\ 2002), 
as well as from thermal gas that surrounds the
nucleus out to a radius of roughly 6 kpc.
Point-like emission is associated with several dozen 
knots in the kpc-scale jet (e.g., Kataoka \et\ 2006)
as well as from a population of X-ray Binaries
(e.g., Kraft \et\ 2000). A soft X-ray spectrum covering the inner few kpc
is thus expected to contain both power-law continuum 
emission from the jet and point-like sources and 
line-like emission associated with the thermal gas. For instance,
Turner \et\ (1997) modeled the soft X-rays with a 0.6 keV thermal plasma component,
a 5 keV component associated with emission from a population 
of X-ray Binaries, and a power-law jet component.

In this paper, we report on an observation of Cen A made with
the {\it Suzaku} observatory in 2005 August.  
The combination of the X-ray Imaging
Spectrometer (XIS) CCD and the Hard X-ray Detector (HXD) instruments 
have yielded a broadband spectrum covering 0.3 to 250 keV,  allowing us
to deconvolve the various broadband emitting and absorbing components in this object.
Furthermore, the exceptional response of the XIS CCD and high signal-to-noise ratio  
of this observation allow us to study narrow emission lines 
in great detail.  This observation has also 
yielded the highest quality soft X-ray spectrum of Cen A obtained so far.
$\S$2 gives a brief overview of the {\it Suzaku} observatory, and describes the 
observation and data reduction. $\S$3 describes the spectral fits.
The results are discussed in $\S$4, and a brief summary is given
in $\S$5.

\section{Observations and Data Reduction}

The nucleus of Cen A was observed by {\it Suzaku} from 2005 August 19 at 03:30 UT until 
August 20 at 09:50 UT, and in fact was the first light target for the HXD.
{\it Suzaku} was launched 2005 July 10 into a low-Earth orbit. It 
has four X-ray telescopes (XRTs; Serlemitsos et al.\ 2007), 
each with a spatial resolution of 2$\arcmin$ (half-power diameter). 
The XRTs focus X-rays onto four X-ray Imaging Spectrometer (XIS; Koyama
et al.\ 2007) CCDs, which are sensitive to 0.2--12 keV X-rays on a 
18$\arcmin$ by 18$\arcmin$
field of view, contain 1024 by 1024 pixel rows each,
and feature an energy resolution of 
$\sim$140 eV at 6 keV. Three CCDs (XIS0, 2, and 3) 
are front-illuminated (FI),
the fourth (XIS1) is back-illuminated (BI) and features 
an enhanced soft X-ray response. 
The XRT/XIS combination yields effective areas per detector of roughly
330 cm$^{2}$ (FI) or 370 cm$^{2}$ (BI) at 1.5 keV,
and 160 cm$^{2}$ (FI) or 110 cm$^{2}$ (BI) at 8 keV.
Each XIS is equipped with two $^{55}$Fe calibration sources
that produce fluorescent Mn K$\alpha$ and K$\beta$ lines and are 
located at the CCD corners.
{\it Suzaku} also features a non-imaging, collimated Hard X-ray Detector
(HXD; Takahashi et al. 2007); its two detectors, PIN and GSO,
combine to yield sensitivity from $\sim$10 to $\sim$700 keV. 
Further details of the {\it Suzaku} observatory are given in
Mitsuda \et\ (2007). 

\subsection{XIS reduction}

The XIS data used in this paper were 
version 0.7 of the screened data (Fujimoto \et\ 2007)
provided by the {\it Suzaku} team. The screening is based on the
following criteria: 
grade 0, 2, 3, 4, and 6 events were used, the {\sc cleansis} script 
was used to remove hot or flickering pixels,
data collected within 256 s of passage through the South Atlantic Anomaly (SAA)
were discarded, and data were selected to be 5$\degr$ in elevation above
the Earth rim (20$\degr$ above the day-Earth rim).
The XIS FI CCDs were in 2x2, 3x3 and 5x5 
editmodes, for a net exposure time after screening of 71.5 (XIS0), 70.3 (XIS2) and 68.6 (XIS3) ks. 
XIS1 was in 3x3 and 5x5 modes, for a net exposure of
68.6 ks.

The source was observed at the nominal center position of the XIS.
The XIS was kept in Normal Mode, but to reduce the risk of photon-pileup, the 
1/4 Window Option was used: each CCD recorded
256 by 1024 pixel rows, with a readout time of 2 s, as opposed to 
8 s when the full 1024 by 1024 window is used.
For each XIS, we extracted a 2$\arcmin$ radius centered on the source.
Spectra were binned to a minimum of 50 counts bin$^{-1}$ to
allow use of the $\chi^2$ statistic.

The count rates observed over the full XIS band were
5.0 ct s$^{-1}$ (average of XIS0, 2 and 3) or 
4.3 ct s$^{-1}$ (XIS BI). These count rates are far below
the limit for photon-pileup in 1/4 Window Mode, $\sim$ 12.5 ct s$^{-1}$.
Furthermore, extracting over annular regions with inner and outer radii
of 0.4$\arcmin$ and 2$\arcmin$, respectively,  yielded spectra identical
in shape to those extracted over 2$\arcmin$ circular regions, indicating
that pile-up was negligible.

Because the soft diffuse emission in Cen A covers nearly the
entire read-out area of the CCD, we used observations
of the north ecliptic pole (NEP) to extract a background spectrum.
{\it Suzaku} observed the NEP on 2005 September 2 to 4.
There was a brief, sharp flare during the NEP observation,
consistent with evidence for solar wind charge exchange emission
(see Fujimoto \et\ 2007). Removing that flare and 
applying the same screening criteria as above yielded a
net exposure time of 93.9 (XIS0), 108.5 (XIS1), 93.3 (XIS2), and 96.6 (XIS3)
ks. We extracted data over a 2$\arcmin$ circle. 
The NEP is known to contain soft X-ray emission lines
(see e.g., Fujimoto \et\ 2007), 
but they are much fainter than those of Cen A (see $\S3$).
The O {\sc VII} and O {\sc VIII} lines in Cen A are a factor of
roughly 10 (20) higher compared to the NEP as seen in the XIS BI
(XIS FIs). Lines near 0.8--0.9 keV such as 
Fe L {\sc XVII} and Ne {\sc IX} in Cen A are typically 
$\sim$60 (100) times brighter in Cen A compared to the NEP
as seen in the XIS BI (XIS FIs).

Average 0.5--2.0 keV count rates for the NEP observation
were  0.004, 0.009, 0.004, and 0.003 
count s$^{-1}$ for XIS0, 1, 2, and 3, respectively;
average 2--10 keV count rates were 0.005, 0.012, 0.005, 
and 0.004 count s$^{-1}$, respectively.

Response matrices and ancillary response files (ARFs) were generated for 
each XIS independently using {\sc xissimrmfgen} and 
{\sc xissimarfgen} version 2006-10-26 (Ishisaki \et\ 2007). The ARF generator takes into 
account the level of hydrocarbon contamination on the optical blocking 
filter. However, the Cen A observation occurred very early in 
the {\it Suzaku} mission, and the level of contamination was quite low: 
we estimate a carbon column density of only $\sim$0.1, 0.2, 0.3 and 
0.5$\times$10$^{18}$ cm$^{-2}$ for XIS0, 1, 2 and 3, respectively.

A consequence of observing with the 1/4 Window Option is that
there are no $^{55}$Fe calibration source lines
on the CCD. This leads to a higher systematic uncertainty in the 
energy scale compared to normal mode (where it
is a few tenths of a percent at most).
To estimate the instrument resolution, we used
the calibration source lines from a 77 ks observation 
of MCG--6-30-15, observed in Normal Mode with no window option
immediately prior to
Cen A. We fit the calibration source spectra with
three Gaussians. Two Gaussians were for the Mn K$\alpha$ doublet
(expected energies 5.899 keV and 5.888 keV), with
energy centroids fixed to be 11 eV apart, and the higher energy
line flux set to twice that of the lower energy one.
The third Gaussian was used to model the K$\beta$ line,
expected at 6.490 keV. We found the average of
all the calibration line widths $\sigma$ to be 9$^{+6}_{-2}$ eV;
residual width may be due e.g., to imperfect CTI correction.

The positional accuracy of {\it Suzaku} is $\sim$1$\arcmin$ at present, as
the spacecraft has been known to exhibit small
attitude variations (``wobble''). 
We checked to make sure that the effect of these attitude variations
was not significant, given that the 
1/4 Window Option was used throughout the observation. 
We generated light curves by extracting over
radii of 
1.5$\arcmin$ and 1.0$\arcmin$ and looked for variations on the 
orbital timescale, but found nothing significant.
  
Average 0.5--2.0 keV net source count rates were 0.24, 0.35, 0.25, and 0.25 
count s$^{-1}$ for XIS0, 1, 2, and 3, respectively.
Average 2--10 keV net source count rates were 4.7, 4.4, 4.5, and 4.5 count s$^{-1}$ for XIS0, 1, 2, and 3, respectively.
Figure 1 shows 2--10 keV light curves for each XIS.
During the observation, the 2--10 keV source flux increased 
by only 10$\%$, with fractional variability amplitudes 
$F_{\rm var}$ (see Vaughan \et\ 2003 for definition) of 
$3.7 \pm 0.2$ $\%$, 
$3.4 \pm 0.2$ $\%$,  
$3.1 \pm 0.2$ $\%$, and 
$3.3 \pm 0.2$ $\%$ for XIS0, 1, 2, and 3, respectively.
The 0.5--2.0 keV flux (determined from model fits; see $\S$3)
was 2.33$\times$10$^{-12}$ erg cm$^{-2}$ s$^{-1}$.
The 2--10 keV flux was 2.12$\times$10$^{-10}$ erg cm$^{-2}$ s$^{-1}$.

\subsection{HXD reduction}

The PIN and GSO source spectra were extracted from cleaned version 1.2 
(pre-1.2-r1) HXD event files provided by the HXD instrument team. We 
first discuss the PIN extraction. Data were selected according to the 
following criteria: at least 500 s since SAA passage, COR $\geq$ 8 GV, 
and day- and night-Earth elevation angles each $\geq$5$\degr$.
Instrumental (non-X-ray) background spectra for the PIN were provided by the 
HXD Team (``background D'' model) generated from a time-dependent 
model.\footnote{Other recently-published {\it Suzaku} results have made use of
the ``background A'' model. However, observations occurring
before 2005 September 2 had a hit-pattern width set to a shorter value
compared to observations occurring after this date. Because applying
the ``background A'' model would yield an underestimate of the true
non-X-ray PIN background, we instead use the ``background D'' model.}
The model utilized the count rate of upper discriminators as the measure of
cosmic ray flux that passed through the silicon PIN diode and yielded
background spectra based on a database of non X-ray background observations
with the PIN (Fukazawa \et\ 2007).  
The systematic uncertainty of the PIN is expected to be $<$5$\%$. 
However, we note that Cen A is one of the brightest AGNs in the PIN
energy range, and the effect of background subtraction error
on the net source spectrum is relatively small.
Both the source and 
background spectra were generated with identical good-time intervals, and 
the exposures were corrected for instrument dead time (a $\sim$5$\%$ 
effect). This yielded a good-time exposure of 60.8 ks.
Data $<$ 12 keV were discarded due to noise contamination
near the lower threshold of the PIN diode. Data above 76 keV were also
discarded: the gain above an internal Bi K$\alpha$ calibration line at 
76 keV is not well-defined, although there are GSO data covering these energies. 
Further details of the HXD in-orbit performance are given in
Kokubun \et\ (2007).
To model the contribution to the total background from the cosmic x-ray 
background (CXB), the spectrum of the form 
9.0$\times$10$^{-9}$(E/3keV)$^{-0.29}$ exp(--E/40keV) erg cm$^{-2}$ s$^{-1}$ sr$^{-1}$ keV$^{-1}$ (Gruber \et\ 1999) was used; the contribution
in the 12--76 keV band was 1.1$\times$10$^{-11}$ erg cm$^{-2}$ s$^{-1}$.

The spectrum was binned to a minimum of 400 count bin$^{-1}$.
We used the response file ae$\_$hxd$\_$pinxinom$\_$20060814.rsp.
The 12--76 keV net source flux and count rate were
7.3$\times$10$^{-10}$ erg cm$^{-2}$ s$^{-1}$ and 1.21 count s$^{-1}$.
The total (X-ray plus particle) background 12--76 keV flux and count rate
were 6.9$\times$10$^{-10}$ erg cm$^{-2}$ s$^{-1}$ and
0.59 count s$^{-1}$.
The orbitally-binned light curve is shown in Figure 1. The light curve
generally increases in flux by $\sim$10--20$\%$ over the observation,
matching the general trend shown by the XIS light curve, although
shorter-timescale differences between the XIS and PIN light curves might be attributed
to uncertainty associated with the PIN background.
Figure 2 shows the net source, background, and total (source + background) spectra.
The source spectrum is always at least 30$\%$ of the total up to $\sim$50 keV.

The HXD GSO data were reduced in a manner similar to the
HXD PIN data. A key contributor to the GSO particle background 
is activation lines, e.g., delayed emission from radioactive 
isotopes induced inside the detector due to interaction with 
SAA particles. To minimize such emission,
data taken within 6000 s of SAA passage were 
discarded; the background is the most reliable and the least variable
during non-SAA orbits. This yielded a good-time exposure of 17.1 ks.
The 'background D' model files provided by the HXD Team were
used (Fukazawa \et\ 2007). The CXB is expected to contribute insignificantly 
to the total GSO background and was ignored. Source and background 
spectra were both binned with {\sc grppha} following recommendations 
from the HXD Team\footnote{group 0 24 25 25 26 2 27 28 2 29 31 3 32 35 4 36 38 3;
group 39 42 4 43 46 4 47 51 5 52 56 5 57 62 6;
group 63 68 6 69 75 7 76 83 8 84 91 8 92 100 9; 
group 101 110 10 111 121 11 122 134 13 135 147 13 148 162 15; 
group 163 178 16 179 196 18 197 216 20 217 238 22 239 262 24;
group 263 288 26 289 317 29 318 349 32 350 384 35 385 422 38;
group 423 465 43 466 511 46}.

The source was detected out to 250 keV; the GSO is sensitive
down to roughly 45 keV. The 45--250 keV net source 
flux and count rate were   
7.2$\times$10$^{-10}$ erg cm$^{-2}$ s$^{-1}$ and 0.83 count s$^{-1}$.
The 45--250 keV background flux and count rate were
1.0$\times$10$^{-8}$ erg cm$^{-2}$ s$^{-1}$ and 9.1 count s$^{-1}$.
Figure 1 shows the GSO light curve.
Figure 2 shows the net source, background, and total (source + background) spectra.
The source spectrum is always at least 5$\%$ of the total below 200 keV.
The response file ae$\_$hxd$\_$gsoxinom$\_$20060321.rmf was used.

Below 100 keV, the GSO field of view is 
34$\arcmin$$\times$34$\arcmin$ FWHM, the same as the PIN.
Above 100 keV, the field of view increases with
photon energy to a maximum of 4.5$\degr$$\times$4.5$\degr$ FWHM
and there is the possibility of source confusion.
The nearest possible contaminating source 
is the blazar MS 1312.1--4221 = GRO J1312--42, located $\sim$2$\degr$ west of Cen A.
According to Steinle et al.\ (1998), the flux of MS 1312.1--4221 in the {\it Compton Gamma-Ray Observatory} ({\it CGRO}) OSSE band
(50--4000 keV) is likely less than a tenth of Cen A's flux. 
MS 1312.1--4221 is much fainter at other X-ray bands as well:
the 1 keV intensity is usually only 1--10$\%$
of the historically observed Cen A intensities (Kinzer \et\ 1995).
We henceforth assume that the contamination from MS 1312.1--4221
is negligible in the GSO band. 

\section{Spectral Analysis}

Given the 2$\arcmin$ half-power diameter of the XRT, the spectrum will contain blended 
contributions from multiple sources.
Previous studies (e.g., Evans \et\ 2004) 
have shown the hard X-rays\footnote{We henceforth denote ``hard X-ray'' emission as 
any intrinsic emission component above 2--3 keV, as opposed to describing emission detected solely with the HXD.}
to be dominated by two power laws;
both hard X-ray components are constrained to lie within a few tenths of an arcsecond ($\sim$5 pc) of the
position of the unresolved radio core
by {\it Chandra} ACIS (Kraft \et\ 2000).
Contributions to the soft X-rays include thermal emission from the diffuse plasma plus
diffuse emission from the kpc-scale jet, with contributions from 
a few of the innermost knots in the kpc-scale jet and other point sources resolved by 
{\it Chandra} ACIS (Kraft \et\ 2000; Kataoka \et\ 2006). Each knot, however, 
is extremely faint in the soft X-rays ($\lesssim$ 6 $\times$ 10$^{-13}$ erg cm$^{-2}$ s$^{-1}$; Feigelson \et\ 1981).

The four XIS spectra were fit separately; 0.4--11.5 keV FI data and
0.3--11.5 keV BI data were included.
We found it reasonable to keep the photon indices tied for all four XISes. while allowing
the relative normalizations for XIS0 and 1 to each vary relative to that for XIS2 and 3,
which were tied together (XIS3/XIS2 was always consistent with, and therefore fixed at, 1.00).
XIS0/XIS2 was typically 1.06; XIS1/XIS2 was typically 0.96
(in our best-fit model, XIS0/XIS2 = 1.065$\pm$0.004 and
XIS1/XIS2 = 0.951$\pm$0.003).
We ignored 1.80--1.87 keV in the XIS FI spectra and 1.75--1.87 keV in 
the XIS BI spectrum due to uncertainties in calibration associated with 
the instrumental Si K edge. This means we cannot directly study the
intrinsic neutral Si K edge in the absorbing material or any  
highly-ionized Si K lines expected 
due to the thermal plasma emission. 
The PIN/XIS2 and GSO/XIS2 normalizations were left free, but were usually 
close to 1.09 and 0.9--1.0, respectively, although results were not strongly 
dependent on these factors (in our best-fit model, PIN/XIS2 = 1.055$\pm$0.022
and GSO/XIS2 = 0.98$\pm$0.06).  
Preliminary results on the relative GSO/PIN normalization
using the Crab suggest GSO/PIN = 0.84 (T.\ Yamasaki et al.\ 2007, in preparation).
Fixing GSO/PIN at 0.84 in the Cen A fits usually resulted in
$\chi^2$ increasing by $\lesssim$5; the photon index
did not change significantly. We left the GSO/PIN normalization free
to attain the lowest $\chi^2$ possible.
All errors on one interesting parameter
correspond to $\Delta\chi^2 = 2.71$ (with all relative
normalizations except XIS3/XIS2 left free). 
Observed spectral features in Cen A are redshifted by 
z = 0.001825 (Graham 1978), the value corresponding to
Cen A's recessional velocity.
To calculate luminosities, a redshift of 0.0009, corresponding to
a luminosity distance of 3.8 Mpc (assuming $H_0 = 70$ km s$^{-1}$ Mpc$^{-1}$)
was used.
Galactic absorption of 8.6$\times$10$^{20}$ cm$^{-2}$ was included.
The abundances of Lodders (2003) were used. Absolute abundances
relative to solar values are denoted by $Z_{\rm Fe}$; a value of 1 
denotes solar abundance.

As shown below, the best model fit includes five overlapping 
broadband components, and changes in how one is fit can potentially 
cause changes in other components. Our strategy in finding the best-fit 
broadband model was to fit the hard band first, then extend it to
the soft band. Relatively more narrow features (lines and edges)
were explored later.

\subsection{Broadband fit}

We first fit the $>$3 keV emission with an absorbed
power law (henceforth denoted PL1) using {\sc zvphabs(cutoffpl)} (Model 1).
The power-law cutoff was kept fixed at 400 keV; 
abundances in the absorber were initially kept fixed at solar.
Residuals are plotted in Figure 3.
The slight hard excesses above $\sim$9 keV in the XIS spectra are
associated with calibration uncertainties known at the time of this writing.
Residuals in the GSO spectrum near 150--200 keV are likely
associated with uncertainties in estimating the strengths of
activation lines in the GSO background (lines at 153 and 196 keV 
are due to $^{153}$Gd and $^{151m}$Eu, respectively);
see Kokubun \et\ (2007) for further details.
GSO residuals near 65 keV are likely associated with 
calibration uncertainties present at the time of this writing.
Large residuals near the expected Fe K$\alpha$ line were apparent.
We then added Fe K$\alpha$ and K$\beta$ lines;
the K$\beta$ energy was fixed at 7.056 keV, its intensity was fixed at
0.13 times that of K$\alpha$, and the widths of the two lines were tied.
The $\chi^2$/$dof$ dropped from 10700/7135 to 7607/7132,
but the residuals in the XIS band
suggested the fit could be improved if we added a second, fainter
absorbed power law (PL2; Model 2). The photon index of the second power law,
$\Gamma_2$, was kept tied to that of PL1, $\Gamma_1$, for simplicity, but
the absorbing column densities $\nh$$_1$ and $\nh$$_2$ were not.
$\chi^2$ dropped by over 140. PL1, with a 1 keV normalization
$\sim$4 times that of PL2, was absorbed by a column
$\nh$$_1$ near 1.0$\times$10$^{23}$ cm$^{-2}$; $\nh$$_2$
was near 5.0$\times$10$^{23}$ cm$^{-2}$. The photon index was
1.773$^{+0.018}_{-0.015}$. Residuals to Model 2 are plotted in Figure 3.


We then included the soft band data, down to 0.3 keV.
Several soft X-ray emission lines are prominent, and are especially
well constrained with XIS1, as illustrated in Figure 4. Simply adding a third power law 
(Model 3) to the soft X-rays thus did not yield an acceptable broadband fit.
Table 1 shows results for the lines when each is fit with a Gaussian; 
we identify them as originating in
O {\sc VII},  O {\sc VIII}, a blend of Fe {\sc XVII} 3s--2p lines (the 
$^3$$P_1$/$^3$$P_2$ doublet and the $^1$$P_1$ line), 
a blend of Fe {\sc XVII} 3d--2p lines ($^1$$P_1$ and $^3$$D_1$),
Ne {\sc IX}, Ne {\sc X}, and Mg {\sc XI}. 
We can use the lines as a check of the absolute energy
scale of the XIS. Assuming that in each He-like triplet, the resonant line is a 
factor of 2 in intensity greater than the forbidden line (roughly consistent with the plasma temperatures
in our best-fit model, below), the measured energy centroids
are within $\lesssim$16 eV of the expected energies. 
In addition, the Fe K$\alpha$ centroid is 6$\pm$3 eV from its expected energy.

The emission lines could be due to gas that is photo- or 
collisionally ionized. A key to distinguishing between the
resulting spectra is that collisional ionization usually leads
to the presence of very strong Fe L emission, such as
a narrow Fe {\sc XVII} emission line near 0.83 keV
and a Fe L emission "bump" near 0.7--0.9 keV.
We modeled the soft emission using {\sc XSTAR} tables appropriate
for emission from photoionized gas, assuming either
one or two zones of photoionized gas. 
Fitting this model always left strong residuals due to improperly modeled
Fe L emission, even when the Fe abundance was left as a free parameter
(in which case values commonly pegged at an upper limit of 10).
The best-fit models assuming either one or two zones   
had $\chi^2$/$dof$ values of 11600/9159 and
11000/9155, respectively.


Attempting to model the soft band emission with a single-temperature
{\sc vapec} component, with all abundances fixed at solar,
also did not yield an acceptable fit. In Model 4,
we tried two {\sc vapec} components, henceforth denoted VAPEC1 and VAPEC2.
The best-fit model had temperatures of
$k_{\rm B}T$ $\sim$ 0.2 keV, to model the He- and H-like O lines,
and $k_{\rm B}T$ $\sim$ 0.6 keV, to model the Fe L and He- and H-like Ne lines. 
However, $\chi^2$/$dof$ was still high, 10395/9160, and there
were still large residuals $<$0.7 keV.

It is plausible that some optically-thin circumnuclear material 
may scatter some of the hard X-ray continuum; a scattered power-law
component was included in the best-fit model of Turner \et\ (1997), 
for instance. In addition to the two {\sc vapec} components, 
we added a third power law (PL3), with photon index $\Gamma_3$ tied to
$\Gamma_1$, assuming that
PL3 undergoes no absorption in excess of the Galactic column. 
Here $\chi^2$/$dof$ was high: 10158/9161. Untying $\Gamma_3$ improved the fit considerably:
$\chi^2$/$dof$ fell to 9806.8/9158; $\Gamma_3$ was 1.16$^{+0.13}_{-0.10}$ (Model 5).

However, it is plausible that PL3 does undergo some absorption
in the host galaxy, so we added a {\sc zvphabs} component to PL3,
and refit with $\Gamma_3$ tied to $\Gamma_1$. 
$\chi^2$/$dof$ fell to 9798.9/9160. The column density of the third
absorber, $\nh$$_3$, was 0.13$\pm$0.06 $\times$10$^{22}$ cm$^{-2}$. The 1 keV normalization of PL3
was 0.8$\%$ of the sum of the normalizations of PL1 and PL2.
Untying $\Gamma_3$ improved the fit considerably:
$\chi^2$/$dof$ fell to 9745.1/9159,
$\Gamma_3$ was 1.28$^{+0.08}_{-0.12}$, and $\nh$$_3$ was $<$0.03 $\times$10$^{22}$ cm$^{-2}$ 
(Model 6). All of the improvement in the fit
when $\Gamma_3$ was thawed came at energies below 2 keV.
PL3 likely represents a blend of scattered emission 
(likely $\lesssim$ 0.8$\%$ of the total nuclear 
hard X-ray continuum) plus X-ray emission associated with the kpc-scale jet, with perhaps some
contribution from XRBs or ULXs in the inner 2 kpc of the host galaxy.
As we have achieved a much better fit with collisional rather than
photoionization models, we will henceforth assume the soft emission lines are collisional in nature, although 
we cannot rule out, e.g., that some portion of the O {\sc VII} and O {\sc VIII}
emission lines may be due to photoionization,  or that there may be a small contribution from 
a O {\sc VII} radiative recombination continuum feature to the observed Fe {\sc XVII} 3s--2p emission complex at 0.732$^{+0.004}_{-0.006}$ keV.
For the moment, we adopt Model 6 as our new baseline broadband model. Best-fit parameters for
Models 5 and 6 are listed in Table 2. Data/model residuals for Models
3, 4, and 6 are shown in Figure 5. 

\subsubsection{Hard X-ray continuum emission}

Evans \et\ (2004) fit the hard X-ray {\it Chandra} and {\it XMM-Newton}
data with a dual power-law model, wherein
$\Gamma_1$ and $\Gamma_2$ differed by about 0.3.
We untied $\Gamma_2$ from $\Gamma_1$. 
However, compared to Model 6, $\chi^2$ only dropped by 1.4,
and the best-fit values of $\Gamma_1$ and $\Gamma_2$
were consistent; the uncertainty on $\Gamma_2$ was $\pm$0.24.
In subsequent fits, we will continue to fix $\Gamma_2$=$\Gamma_1$.
 
There has been no strong evidence from past observations for a
strong Compton reflection hump in Cen A. We added a Compton
reflection component to Model 6 using {\sc PEXRAV},
keeping the input normalization equal to that of PL 1, 
the inclination fixed at 30$\degr$, the high-energy cutoff fixed at 400 keV,
assuming solar abundances, and leaving the PIN/XIS2 normalization free.
The best fit-model preferred a value of the reflection fraction $R$
(defined as $\Omega$/2$\pi$, where $\Omega$ is the solid
angle subtended by the reflector) of 0, with
an upper limit of 0.05, consistent with
previous suggestions that the Fe K$\alpha$ line originates in
Compton-thin material. This limit is identical to that obtained
by Benlloch \et\ (2001) using {\it Rossi X-ray Timing Explorer} ({\it RXTE}) 
data. Contour plots of $R$ versus $\Gamma_1$ and versus 
the PIN/XIS2 relative instrument normalization are shown in 
Figure 6.

Using {\it CGRO} OSSE, 
Kinzer \et\ (1995) found evidence for a high-energy
cutoff which appeared at 300 keV to $\sim$700 keV in
the highest- to lowest-flux states, respectively.
In addition, Steinle \et\ (1998) fitted data combined from
{\it CGRO} OSSE, COMPTEL and EGRET;
one of the spectral breaks they found occurred 
near 150 keV, with the spectral index $\alpha$ changing from
1.74$^{+0.05}_{-0.06}$ below the break to
2.3$\pm$0.1 above it.
Keeping the cutoff energies for PL1 and PL2 in Model 6
equal to each other, we found a lower limit of 400 keV to any
break in the power law.
Future improvements to background modeling of the GSO
may eventually allow us to establish more strict constraints.

\subsection{Hard X-ray emission lines and absorption edges}

In addition to that from Fe, fluorescent K$\alpha$ lines from
Si and S have been claimed by Sugizaki \et\ (1997) and Evans \et\ (2004).
We added five more Gaussians, each with width tied
to that of the Fe K$\alpha$ line; it was significant
at $>$90$\%$ confidence according to the $F$-test to add each line.
The centroid energies are consistent with K$\alpha$ emission from
Si, S, Ca, Ar, and Ni; detection of the latter three lines are
reported here for the first time. 
The decrease in $\chi^2$ for each line, including the 
Fe K$\alpha$ and K$\beta$ lines,
along with energy centroids, fluxes, and observed equivalent widths
$EW_{\rm obs}$, are listed in Table 3.
In the case of Si, we caution that since data near 1.8 keV have been ignored due to  
calibration uncertainties, 
values of flux and equivalent width may be uncertain
(e.g., some fraction of the large value of $\Delta\chi^2$ may be
due to calibration uncertainties).
The detection of the Ni line is robust:
there does exist a Ni K$\alpha$ line in the NEP background
spectrum, but it is 200 times fainter than the source spectrum line.
Data/model residuals to models with all six K$\alpha$ lines
and the Fe K$\beta$ line removed are shown in Figures 7--9.

The Fe K$\alpha$ line has a best-fit measured width of 14 eV, 
although formally the value is
only an upper limit, 24 eV. The intrinsic width of the line is then found
by subtracting in quadrature
the $^{55}$Fe calibration line width of 9$^{+6}_{-2}$ eV
from the measured line width. We infer an intrinsic line 
width of $<$ 23 eV, which corresponds to a FWHM velocity of 
$<$2500 km s$^{-1}$.

We next explored the abundances of the 
hard X-ray absorbers (Model 7). The K-shell edges for
Fe, Ca, and S are detected at $>$99.999$\%$ confidence, determined by
setting each {\sc zvphabs} abundance value to 0 and refitting. 
The detection of the Ca K edge is likely robust: the response of a line at 6.4 keV
includes a Si escape feature near 4.5 keV, but it is a factor of 
more than 50 fainter
than the continuum and likely does not influence the Ca edge.
The intrinsically weak Ar K and Ni K edges are not detected 
at high confidence; in the latter case, the
decrease in the effective area of the XIS 
above 8 keV is also a factor.
We re-fit the data, allowing $Z_{\rm Fe}$,
$Z_{\rm Ca}$, and $Z_{\rm S}$ to vary,
and keeping the abundances of all three absorbers
tied to each other for simplicity.
We also kept all other abundances fixed at solar.
Best-fit model parameters for Model 7 are listed in Table 2; the edges are also
displayed in Figures 7--9.
The abundance results are listed in Table 4; note that the Fe abundance
is inconsistent with solar at 3$\sigma$ confidence.
Given the column density of the absorbing material and the 2--3 keV rollover,
most of the rollover is the result of absorption by K-shell O
(with some contributions to the total opacity from
K-shell Ne and L-shell Fe). We are thus actually measuring
abundances relative primarily to O, not to H. The abundances listed
relative to H in Table 4 implicitly assume $Z_{\rm O}$ = 1.0.

To find the equivalent optical depth for each edge, we 
re-fit the data with $Z_{\rm Fe}$,
$Z_{\rm Ca}$, and $Z_{\rm S}$ set to zero in the {\sc zvphabs} components, but with
edges at the appropriate energies. The results, listed in Table 4, show that
the energy of each edge is consistent with absorption by neutral atoms.

The Fe K$\beta$/K$\alpha$ intensity ratio can yield insight into the
ionization state of the line-emitting gas. However, in the present spectrum,
the K$\beta$ line is somewhat blended with the Fe K edge.
However, we note that the Fe K edge energy and depth are robust to the properties of
the K$\beta$ line, and the energy can be used to constrain the ionization state
of the material, as discussed further in $\S$4.

We searched for any possible Compton shoulder to the Fe K$\alpha$ line by 
adding a Gaussian near 6.24 keV, but there was no improvement in the fit.
We find an upper limit of 1.7$\times$10$^{-5}$ ph cm$^{-2}$ s$^{-1}$
to Compton shoulder emission ($EW$ $<$ 10 eV, determined relative to a
locally-fit continuum).

Grandi \et\ (2003) claimed emission due to ionized Fe
near 6.8 keV using {\it BeppoSAX}. We added a narrow 
Gaussian at the rest energies of Fe {\sc XXV} and Fe {\sc XXVI}, 
with no change in the fit statistic in either case;
upper limits to the $EW$ of emission lines were 3 and 1 eV, respectively.

\subsection{Abundances of the thermal plasma}

We next explored the abundances $Z$ 
of the thermal plasma relative to solar values.
Abundances for the VAPEC1 and VAPEC2 components were kept tied
to each other. We thawed one element at a time from solar, while
assuming solar abundances for all other elements.
Abundances relative to H can be measured accurately
if the continuum is well-determined and the thermal emission
is not too heavily dominated by lines.
Given the presence of PL3, one must keep in
mind that there could be systematic effects
associated with some derived abundance values. 

We did not find any strong evidence for non-solar abundances for
C, N, or O. The best chance to constrain $Z_{\rm C}$ in our model is via the
C {\sc VI} Ly $\alpha$
and Ly $\beta$ lines at 0.368 and 0.436 keV, respectively.
However, there are very few counts at these energies, and the
PL3 component dominates at these energies. 
Similarly, $Z_{\rm N}$ could, in principle, be constrained via the
N {\sc VII} Ly$\alpha$ line, but the line is not robustly detected. 
Allowing $Z_{\rm O}$ to vary resulted in a best-fit value of
0.95$\pm$0.35 solar, with $\chi^2$ decreasing by only 1.7.
We assume $Z_{\rm C}$, $Z_{\rm N}$ and $Z_{\rm O}$ are solar 
for the rest of the paper.

We did not test the Si abundance: Si {\sc XIII} emission cannot be
constrained, as we have ignored data near the Si K instrument edge.
The best-fit model does predict a modest Si {\sc XIV} emission line
at 2.05 keV, but at that energy, the PL3 component and the
absorbed hard X-ray power laws dominate.

In addition to Fe {\sc XVII} emission lines,
Fe is also responsible for a large bump in the continuum emission
from about 0.7 to 1.0 keV. Adjusting the Fe abundance
causes the PL3 component to vary in normalization to
compensate, creating systematic uncertainty 
to $Z_{\rm Fe}$ derived from L-shell emission.
We henceforth rely on the 
abundance determined using the K-edge depth ($\S$3.3). 

We did, however, find significant improvements to the fit when the
Ne or Mg abundances were varied.
In the case of Ne, the best-fit abundance was 2.7$^{+0.3}_{-0.2}$ solar;
$\chi^2$ decreased by 71 ($F$-test probability 2$\times$10$^{-15}$).
For Mg, the best-fit abundance was 1.6$\pm$0.3 solar,
with $\chi^2$ decreasing by 19 ($F$-test probability 2$\times$10$^{-5}$).

We re-fit the data with the Ne and Mg abundances in VAPEC1 and VAPEC2 thawed
(Model 8).
We left the {\sc vapec} abundances for C, N, O, Si, and Fe fixed at solar (fixing
$Z_{\rm Fe}$ at 1.17, the value derived from the Fe K edge, had minimal impact).
In the best-fit model, $Z_{\rm Ne}$ and 
$Z_{\rm Mg}$ were 2.7$\pm$0.3, and 1.9$\pm$0.4, 
respectively. The temperature of the VAPEC2 component fell slightly to 
0.56$^{+0.01}_{-0.03}$ keV.
With the exception of these parameters, all other previously-derived 
parameters did not change significantly compared to Model 7.
The best-fit parameters for Model 8 are listed in Table 2.
A contour plot of $\nh$$_1$ versus the photon index and normalization
of the primary power law is shown in Figure 10.
An unfolded model spectrum is plotted in Figure 11.

 
\subsection{Continuum variability}

Having decomposed the broadband spectrum, it is natural to ask
whether the observed 2--10 keV variability (see Figure 1)
is due to variations in the flux of PL1, PL2, or both.
Figure 12 shows the 2--4 and 5--10 keV light curves,
summed over all four XISes and normalized by their respective means; 
the 2--4 keV band is dominated by PL1; 
PL3, a component that is not expected to display short-term variability,
is an average factor of 30 times fainter in this band.
In the 5--10 keV band, PL1 is a factor of roughly 6 times 
brighter than PL2. 
Fractional variability amplitudes for these two
bands are similar: $F_{\rm var,2-4}$ = 3.0 $\pm$ 0.2$\%$ and
$F_{\rm var,5-10}$ = 3.5 $\pm$ 0.1 $\%$.
Moreover, one can see that the two mean-normalized light curves display
virtually identical variability trends.

The fact that the 2--4 keV band varies implies that PL1 does vary
in flux. If PL2 is constant or only negligibly variable, 
then variability in the 5--10 keV should be diluted
relative to that in the 2--4 keV band\footnote{If a lightcurve $C$ consists of 
a variable component $A$ plus a constant component $B$,
then $F_{\rm var} (C)$ will be equal to 
$F_{\rm var} (A) / (1 + \frac{\mu_B}{\mu_A})$, 
where $\mu_A$ and $\mu_B$ are the average count rates of lightcurves 
$A$ and $B$, respectively. For instance, in the case of Cen A, if PL1
is an average of 6 times brighter than PL2 in the 5--10 keV band, 
and PL2 is constant,
than $F_{\rm var,5-10}$ should be $6/7 \times F_{\rm var}$(PL1) 
(assuming emission from PL3 is negligible).}.
However, it is not the case that
$F_{\rm var,5-10}$ is significantly lower than $F_{\rm var,2-4}$;
in fact, $F_{\rm var,5-10}$ is slightly higher than $F_{\rm var,2-4}$.
On the other hand, with such small $F_{\rm var}$ values 
and the fact that PL1 is so much brighter than PL2, it is 
difficult to accurately conclude anything about whether PL2 is constant,
as variable, or more variable than PL1.

While we have confirmed that PL1 varies,
the similarity in $F_{\rm var}$ for both bands
and the similarity in variability trends
means we cannot rule out the idea that both PL1 and PL2
vary in concert, as expected if a partial-covering scenario applies.
Further progress can be made through continued broadband monitoring
spanning a larger flux range.

\section{Discussion}

The broad bandpass of the XIS and HXD aboard {\it Suzaku} has allowed us to obtain
a high-quality spectrum spanning nearly 3 decades in photon energy.
The long exposure time, narrow response,
and the low background make the XIS spectrum
one of the highest quality CCD spectra of Cen A to
date, particularly below 2 keV.

In $\S$4.1, we compare our results to other recent works to
test for evidence of long-term variations in several observed parameters.
The dual power laws that dominate the hard X-rays, and the material that absorbs them,
are both discussed in $\S$4.2. That same absorbing material is likely the origin of the
fluorescent emission lines; the nature of the Fe K$\alpha$ line is discussed in $\S$4.3.
The soft X-ray emission from the thermal plasma is discussed in $\S$4.4. 
Finally, in $\S$4.5, element abundances are derived from the hard X-ray absorption edges 
and fluorescent emission lines, the first time abundances in the absorbing gas of Cen A
have been measured. In conjunction with abundances measured using the
thermal plasma lines, 
we show that the abundances are consistent with enrichment from star formation.

\subsection{Comparison to previous observations}

We can compare our results to previous works to search for any variations in flux, photon index, 
column densities, and Fe K$\alpha$ line flux. 

We find an absorbed 2--10 keV flux of 2.12$\times$10$^{-10}$ erg cm$^{-2}$ s$^{-1}$,
similar to values obtained by Rothschild \et\ (2006) for {\it RXTE} and {\it INTEGRAL}
observations of Cen A between 2003 and 2004. We find a 20--100 keV flux of 6.4$\times$10$^{-10}$ 
erg cm$^{-2}$ s$^{-1}$, within the range of 20--100 keV fluxes measured by Rothschild \et\ (2006).
We measure an unabsorbed 4--7 keV continuum flux of 1.5$\times$10$^{-10}$ 
erg cm$^{-2}$ s$^{-1}$ and a 2--10 keV absorbed luminosity of 
2.9$\times$10$^{41}$ erg s$^{-1}$, values almost identical to those 
found by Evans \et\ (2004) during the 2001 {\it XMM-Newton} observation.
These facts suggest the source has not varied in hard X-ray flux greatly over 
the last 5 years.

The photon index, power-law normalization, and absorbing column density
of the primary (brighter) power law, PL1, are all similar to values obtained
by Rothschild \et\ (2006). Evans \et\ (2004), modeling
{\it XMM-Newton} observations in 2001 and 2002 and a {\it Chandra}
observation in 2002, fit partial-covering models ($\Gamma_1$=$\Gamma_2$) and
dual power-law models ($\Gamma_2$ not tied to $\Gamma_1$). In all previous observations, the
parameters of the primary power law and its absorber
are very similar to those for the {\it Suzaku} observation, suggesting
a lack of strong variations over the last 5 years. 

The $EW_{\rm obs}$ of the Fe K$\alpha$ line in the {\it Suzaku} observation
is consistent with that measured from the {\it XMM-Newton} observations;
we measure the Fe K$\alpha$ line intensity to be
2.32 $\pm$ 0.1 $\times$ 10$^{-4}$ ph cm$^{-2}$ s$^{-1}$ and the
unabsorbed Fe K$\alpha$ flux to be 2.6$\times$10$^{-12}$ erg
cm$^{-2}$ s$^{-1}$, consistent with the {\it Chandra} and 2001 {\it XMM-Newton}
observations (Evans \et\ 2004). 
Our $EW_{\rm obs}$ values for the Si and S lines are consistent with
those obtained using {\it Chandra} HETGS.
The {\it Suzaku} observation thus shows no evidence for variability of
the Fe, Si, or S K$\alpha$ lines compared to the 2001--2002 observations.

\subsection{Continuum emission components}

We now turn our attention to the nature of the two hard X-ray power laws, PL1 and PL2, and the less-absorbed, 
soft X-ray power law PL3. The absorbers for PL1 and PL2 are similar in column density to absorbers seen in some 
other radio galaxies (Wozniak \et\ 1998). Cen A's dust lane provides an optical extinction of 3--6 mag, which, 
assuming a Galactic gas/dust ratio, corresponds to a column $N_{\rm H}$ $\sim$ 5--10 $\times$10$^{21}$ cm$^{-2}$. 
The circumnuclear material within a few hundred pc of the black hole, however, is believed to have a much higher 
column, $\sim$10$^{23}$ cm$^{-2}$. Both hard X-ray absorbers are thus more likely associated with the circumnuclear 
material than the optical dust lane; PL3, meanwhile, is consistent with absorption by the optical dust lane only.

Historically, it has not been clear whether the X-ray emission is associated with jet or accretion processes. As far 
as jet emission is concerned, several papers in the 1980s (e.g., Burns \et\ 1983) suggested that the X-ray emission 
was synchrotron emission. However, by constructing the broadband SED, Chiaberge \et\ (2001) concluded that the X-ray 
continuum was primarily inverse Compton emission, with the synchrotron component peaking in the far-IR and falling by 
10$^{15-16}$ Hz. We can gauge whether one or more of the three power laws observed by {\it Suzaku} are consistent with X-ray 
emission from a jet by using the radio--X-ray luminosity density correlations of Canosa \et\ (1999) and Evans \et\ (2006), 
which are based on samples of low-redshift radio galaxies. The total 5 GHz luminosity density of the nucleus of Cen A is 
$\sim$6$\times$10$^{20}$ W Hz$^{-1}$ sr$^{-1}$ (e.g., Evans \et\ 2006). This value corresponds to a 1 keV luminosity density 
of $\sim$1--2$\times$10$^{14}$ W Hz$^{-1}$ sr$^{-1}$. The 1 keV power-law normalizations for PL1, PL2, and PL3 correspond to, 
respectively, 1 keV flux densities of 70, 12 and 0.4 $\mu$Jy, or 1 keV luminosity densities of $\sim$ 8$\times$10$^{15}$, 
1$\times$10$^{15}$, and 4$\times$10$^{13}$ W Hz$^{-1}$ sr$^{-1}$. It is therefore plausible that PL3 could correspond to 
jet emission. The X-ray luminosity density of PL2 places it an order of magnitude above the Canosa \et\ (1999) and Evans 
\et\ (2006) relations; however, given that there is roughly an order of magnitude of scatter in the relations, it is still 
possible that PL2 could be associated with jet emission,

We can now revisit previous suggestions that there could exist two physically distinct sites of hard X-ray continuum emission.
Our best-fit model indicates components with unabsorbed 2--10 keV luminosities of 6.7$\times$10$^{41}$ erg s$^{-1}$ (PL1) and 
1.3$\times$10$^{41}$ erg s$^{-1}$ (PL2), each independently absorbed by different columns. Evans \et\ (2004) suggested that 
the primary power law, PL1, arises via Bondi accretion onto the black hole, consistent with the low accretion rate of Cen A 
(0.2$\%$ of Eddington), and consistent with PL1 representing X-ray emission far in excess above what is expected from jet emission.
Evans \et\ (2004) also suggested that their secondary power-law component could be associated with X-ray emission from the 
pc-scale jet. Our PL2 component has a 1 keV normalization $\gtrsim$ 3 times higher and an absorbing column density a factor 
of $\gtrsim$10 higher compared to the secondary power-law component detected by Evans \et\ (2004). However, a one-to-one 
correspondence between our PL2 and the {\it Chandra}/{\it XMM-Newton} secondary power law is highly uncertain. We are 
using larger PSFs and extraction radii than Evans \et\ (2004); the secondary power-law component detected by Evans \et\ (2004)  
could be overwhelmed by non-nuclear emission (e.g., what we detect as PL3) in the {\it Suzaku} spectrum. Meanwhile, 
{\it Chandra} HETGS's low effective area $>$ 5 keV makes it relatively insensitive to the presence of a power-law component
absorbed by a column 7$\times$10$^{23}$ cm$^{-2}$.  It is therefore possible, though not certain, that PL2 could correspond 
to X-ray emission associated with the pc-scale jet. In the soft X-rays, PL3 could be the sum of diffuse X-ray emission and 
knots within the kpc scale jet, scattered nuclear emission, and emission from other point sources resolved by {\it Chandra} ACIS. 
The 0.5--3 keV luminosity of PL3 is $ 1 \times 10^{40}$ erg s$^{-1}$, similar to the sum of the jet and knot component 
luminosities as reported by Kraft \et\ (2002) and Kataoka \et\ (2006). PL3 may additionally reflect uncertainties in modeling 
the thermal gas. 

Alternatively, as suggested by Turner \et\ (1997), a partial-covering model, with only one site of 
hard X-ray emission, is applicable, as the photon indices of PL1 and PL2 are consistent.
Our best-fit model is consistent with the idea that the central hard X-ray source is powered by accretion.
84$\%$ of the sky as seen from the central hard X-ray source
is covered by a column 1.5$\times$10$^{23}$ cm$^{-2}$,
and the remaining 16$\%$ by a column 7$\times$10$^{23}$ cm$^{-2}$. In this model, PL3 
represents the sum of X-ray emission from both the kpc-scale and pc-scale jets, any
scattered nuclear emission, jet knots, XRBs and other point sources, all modified by dust lane absorption,
along with any uncertainty in modeling the thermal gas. 

Finally, we comment on the absence of a strong Compton reflection component.
The low $R$ value could of course indicate the absence of a large amount of 
Compton-thick material that is brightly illuminated by the primary continuum
and also efficiently radiates the reflection continuum. 
An alternate suggestion is that the Compton hump is
diluted by a power-law component, e.g., from one of the
jets. However, the large Fe line flux observed is inconsistent with
this notion. In addition, Wozniak \et\ (1998) demonstrated that
spectral fitting of several radio galaxies
that also displayed weak or non-existent Compton humps
is inconsistent with this notion.

\subsection{The origin of the Fe K$\alpha$ line}

Assuming that the line originates in gas that is in virialized
orbit around the black hole, we can
estimate the distance $r$ from the black hole to the Fe K$\alpha$ 
line-emitting gas. The width of the Fe K$\alpha$ line corresponds to a
FWHM velocity $v_{\rm FWHM}$ of $<$2500 km s$^{-1}$.
Assuming that the velocity dispersion is related to
$v_{\rm FWHM}$ as $<$$v^2$$>$ = $\frac{3}{4}v^2_{\rm FWHM}$ (Netzer \et\ 1990),
assuming a black hole mass $M_{\rm BH}$ of 
2$\times$10$^8$ $\Msun$, and using
$G$$M_{\rm BH}$ = $r$$v^2$, we find a lower limit to $r$ of 
6$\times$10$^{15}$ m = 200 light-days.
The best-fit edge energy, 7.103$\pm$0.015 keV, is consistent with absorption by Fe {\sc I},
although ionization stages up to Fe $\sim${\sc V} cannot be ruled out (Kallman \et\ 2004).

It is plausible that the same material that absorbs the hard X-rays
is responsible for producing the fluorescent lines.
The fact that no Compton hump or 6.2 keV Compton shoulder
are seen, indicating an origin for the Fe K$\alpha$ line in Compton-thin
material, supports this notion. 
On the other hand, the material cannot have a column substantially less than
10$^{\sim 22}$ cm$^{-2}$ or else there would be insufficient optical depth
to produce a prominent Fe K line.
As an estimate of the Fe K$\alpha$ equivalent width expected in this case, we can use the following equation:
\begin{equation}
EW_{\rm calc} = f_{\rm c} \omega f_{\rm K\alpha} A \frac{\int^{\infty}_{E_{\rm K edge}}P(E) \sigma_{\rm ph}(E) N_{\rm H} dE}{P(E_{\rm line})}
\end{equation}
This method assumes an origin in optically-thin gas that completely surrounds a 
single X-ray continuum source and is uniform in column density. 
Emission is assumed to be isotropic.
Here, $f_{\rm c}$ is the covering fraction, initially assumed to be 1.0.
$\omega$ is the fluorescent yield: the value for Fe, 0.34, was taken from 
Kallman \et\ (2004). $f_{\rm K\alpha}$ is the fraction of photons that go into the K$\alpha$ line
as opposed to the K$\beta$ line; this is 0.89 for Fe {\sc I}.
$A$ is the number abundance relative to hydrogen. 
Initially, we assumed solar abundances, using Lodders (2003).
$P(E)$ is the spectrum of the illuminating continuum at energy $E$;
$E_{\rm line}$ is the K$\alpha$ emission line energy.
The illuminating continuum is assumed to be the sum of both unabsorbed hard X-ray
power-law components in Cen A, as $EW_{\rm obs}$ values were determined relative to
the total, local continuum. 
We note that the fainter power-law component cannot reproduce the
entire Fe line as $EW_{\rm obs}$ is too large; the primary power law
must be responsible for the bulk of the line. Finally, 
$\sigma_{\rm ph}(E)$ is the photoionization cross section 
assuming absorption by K-shell electrons only; all cross sections were taken
from Veigele (1973\footnote{http://www.pa.uky.edu/$\sim$verner/photo.html}).
$N_{\rm H}$ is the column density. The two hard X-ray absorbing components 
in Model 8 have column densities 1.5 and 7 $\times$ 10$^{23}$ cm$^{-2}$, which attenuate 
power laws with normalizations in a 5.3:1 ratio. We therefore use a weighted average 
$N_{\rm H}$ column of 2.2 $\times$ 10$^{23}$ cm$^{-2}$.
(Strictly speaking, 
the observation in Cen A of two different column densities argues against
the use of a single, uniform column; however, this weighted average suffices for our purpose
of estimating the total Fe K$\alpha$ $EW$ expected from both absorbers.)

The value of $EW_{\rm calc}$ is 128 eV, a factor of 1.5 higher than
$EW_{\rm obs}$.  Assuming that $Z_{\rm Fe}$ = 1.17, 
as the Fe K absorption edge depth suggests, yields $EW_{\rm calc}$ = 153 eV.
Possible explanations for this discrepancy include the following:
(1) The initial assumption of a 100$\%$ covering fraction for the line-emitting
gas may be incorrect. Assuming isotropic emission, $f_{\rm c}$ = 0.7 ($Z_{\rm Fe}$ = 1.0) or
0.6 ($Z_{\rm Fe}$=1.17) may fit the data. Such values would be more consistent with
an origin in a disk as opposed to a shell completely surrounding the core.
(2) The illuminating X-ray continuum may be anisotropic. For instance, 
Rothschild \et\ (2006) found that $I_{\rm K\alpha}$ was not correlated
with $N_{\rm H}$ and suggested that the line-emitting gas
may lie along the VLBI jet some distance from the black hole.
(3) The line-emitting gas may be responding to an illuminating flux that 
is $\sim$0.6 of the observed flux. This is possible, for instance, if
the line-emitting gas is $\gtrsim$200 light-days from the X-ray continuum origin,
and does not lie on our line of sight to the continuum origin; the observed line flux
is a response to the continuum flux convolved with a $>$200-day time delay.
{\it Swift}/BAT monitoring indicates that the average 14--195 keV flux of Cen A in early 2005
was very roughly 15$\%$ higher compared to the flux in 2005 August (private communcation from the {\it Swift} team in 2007\footnote{See also
the {\it Swift}/BAT Transient Monitoring web page, http://swift.gsfc.nasa.gov/docs/swift/results/transients/}),
possibly explaining some of the line flux discrepancy.
Continued long-term monitoring of the continuum and line fluxes may shed light on this issue.
Alternatively, the line of sight from the continuum origin to the line-emitting
gas may be blocked, by a clump of gas 
that absorbs 40$\%$ of the $>$7 keV flux and that 
does not lie along our line of sight to the continuum.
(4) It is plausible that our initial assumption of a uniform-column density absorber is incorrect and
the absorbers are instead clumpy; this is 
more consistent with the observation of two different columns.
There could be higher columns lying out of our line of sight
to the X-ray continuum source. For instance, 
we can use Eq.\ 1 of Wozniak \et\ (1998), which gives the expected Fe K$\alpha$ intensity assuming a cloud 
with a column $\gtrsim$ 10$^{23}$ cm$^{-2}$ lying off the line of sight and subtending
a fraction $\Omega/4\pi$ of the sky as seen from the source, as a function of the 
K edge optical depth and the spectral index and normalization of the illuminating continuum.
We find that a solid angle of $\Omega/4\pi$ $\sim$ 0.5 yields a predicted
line intensity compatible with what {\it Suzaku} observes.
Such a covering fraction suggests a torus or some sort of infinite slab, e.g., a 
disk-like structure, although the current data cannot constrain whether
such a disk would be in the form of a ``standard'' optically-thick, geometrically-thin
disk at all radii or a ``disk + sphere'' hybrid (e.g., Esin, McClintock \& 
Narayan 1997), wherein the inner parts of a standard thin disk are replaced by 
a radiatively-inefficient flow such as an advection dominated accretion flow (Narayan \& Yi 1995).


\subsection{Soft X-ray emission from the thermal plasma}

The {\it Suzaku} XIS has yielded the best
soft X-ray spectrum obtained to date for Cen A, revealing 
emission lines that likely originate in the thermal plasma
that extends from the nucleus to a radius of $\sim$6 kpc.
At least the inner parts of the gas are likely heated by the AGN
central engine and/or the nuclear starburst;
O'Sullivan, Ponman \& Collins (2003) demonstrated
that the gas in the central 2 kpc of Cen A has a much higher 
temperature then the $>$2 to 14 kpc gas in the halo.

Photoionization does not yield as good a fit to the
{\it Suzaku} data as collisional ionization models, 
as the former cannot reproduce the strong observed Fe L emission
complex. Specifically, to obtain a good 
model fit, we have assumed that the emission is entirely collisional in nature,
and used  a two-temperature
{\sc VAPEC} model with $k_{\rm B}T$ near 0.2 and 0.6 keV.
This is similar to results obtained in a wide variety of
spiral and starburst galaxies (see references in Strickland \et\ 2004).
However, we are likely sampling
thermal emission from plasma spanning a range of temperatures.
For example, the soft X-rays are likely 
not completely dominated by emission from gas with $k_{\rm B}T$
above 0.6 keV, or else we would detect 
a H-like Mg line at least as intense as the He-like Mg line,
and the He-like Ne line would be extremely faint.
Similarly, emission from gas with $k_{\rm B}T$ $<$ 0.20 keV
does not dominate, as the Ne {\sc X} would be extremely faint.

\subsection{Constraints on abundances}

We have derived estimates of $Z_{\rm S}$, $Z_{\rm Ca}$, and $Z_{\rm Fe}$ using the
K shell edge depths, and estimates of $Z_{\rm O}$, $Z_{\rm Ne}$, and $Z_{\rm Mg}$
from the thermal plasma (for the remainder of this paper,
we assume that the abundances in the extended thermal plasma
and in the circumnuclear absorbing material are identical).  

We can also estimate abundances relative to Fe, e.g., $Z_{\rm Si}$/$Z_{\rm Fe}$,
from the K$\alpha$ fluorescent lines and Eq.\ 1.
Values of $\omega$ were taken from the X-ray Data Booklet\footnote{http://xdb.lbl.gov}.
For Ni and Ca,  $f_{\rm K\alpha}$ = 0.89 and 0.88, respectively, assuming neutral atoms (Bambynek \et\ 1972).
We assume $f_{\rm K\alpha}$ = 0.9 for all other elements.
The values of $EW_{\rm calc}$, calculated assuming solar abundances and a covering fraction $f_{\rm c}$=1.0,
are listed in Table 3. 

Relative abundances are calculated as, e.g., $Z_{\rm Si}$/$Z_{\rm Fe}$ = 
($EW_{\rm obs, Si}$/$EW_{\rm calc, Si}$) /
($EW_{\rm obs, Fe}$/$EW_{\rm calc, Fe}$), where
$EW_{\rm calc, Fe}$ has been estimated using the 
abundance derived from the Fe K edge depth, $Z_{\rm Fe}$=1.17, and
the solar abundance was used for $EW_{\rm calc, Si}$. 
We caution that 
for Ar, Ca, and Ni, the lines are weak and the statistics are poor. 
For Si, there may be systematic effects
due to current XIS calibration as well as from the fact that
the continuum near 1.7 keV is a mixture of all three power-law components.
Table 5 lists the abundance ratios. There is agreement on the values of
$Z_{\rm S}$/$Z_{\rm Fe}$ derived from the edges and from the lines.

We observe [m/H] = log($Z_{\rm Fe}$) = +0.1, a slightly higher metallicity 
value than in the metal-rich population of globular clusters at 2--20 kpc radii
([m/H] $\sim -0.1$, Peng \et\ 2004) or in the outer stellar bulge component 
(e.g., [m/H] $\sim$ --0.2 at a radius of 8 kpc, Harris \& Harris 2002).
There have been previous indications of high metallicities
in the central regions of Cen A, i.e., indications of 
overabundances of N and O in young H {\sc II} regions in the 
inner few kpc (Moellenhoff 1981).
Our observed [m/H] value is only slightly 
higher than values for the ISM of other 
early-type galaxies (Humphrey \& Buote 2006).
It is similar to values for the stellar components of large ellipticals
in general, [m/H] $\sim 0.0-0.4$ 
(Henry \& Worthey 1999). Trager \et\ (2000) also found the stellar populations of
eight field ellipticals to have [m/H] $\sim 0.0-0.4$, though it may not be straightforward
to compare Cen A to field ellipticals due to Cen A's recent merger.
Our observed value of $Z_{\rm Mg}/Z_{\rm Fe}$ is consistent with the 
value of 2 obtained by Peng \et\ (2004) for Cen A's globular clusters. 

However, our observed [m/H] value is much higher than the 
average value (of a very wide distribution) found in the halo stars of Cen A,
[m/H] = --0.4 at radii of 20--30 kpc (Harris \& Harris 2000,
Harris, Harris \& Poole 1999).  
It is therefore unlikely that the circumnuclear material and
the diffuse plasma have directly originated in the relatively metal-poor outer halo,
unless enrichment via local star formation has occurred.

The premerger origin of the circumnuclear material cannot be known for certain,
although it is plausible that during the merger, gaseous material
originally in the outer portions of the progenitor galaxies
lost sufficient angular momentum to be transported to the nuclear regions
(Toomre \& Toomre 1972). 
Furthermore, connections between the  
nuclear gas and starburst activity at radii of hundreds of pc and less are well-known
(e.g., Hopkins \et\ 2006 and references therein). The likelihood that Cen A's
circumnuclear gas has been enriched due to starburst processes is therefore high.

The degree of metallicity enrichment depends on many factors,
including the IMF, star formation rate, and the total gas consumed,
but simulations of disk-disk mergers incorporating star formation
by Mihos \&  Hernquist (1994)
demonstrated that [m/H] values of 0.3--0.4 at the center of 
the resulting merged ellipticals are plausible.
The metallicity value derived from the {\it Suzaku}
data is consistent with enrichment due to the ongoing 
star formation in the inner bulge,
as a result of the merging process.

The observed relative element abundances also support enrichment.
We calculated the expected relative abundances due to enrichment by
Type II and Type Ia SNe, ignoring enrichment via stellar wind processes.
We used the abundances relative to solar as found in 
Tsujimoto \et\ (1995; their Figures 1--2),
assuming a Salpeter IMF, converting to the abundances of Lodders (2003), and
assuming one Type Ia explosion for every 10 Type II explosions. 
For the ratios listed in Table 5, 
we expect $Z_{\rm O}/Z_{\rm Fe}$ $\sim$2.5,
$Z_{\rm Ne}/Z_{\rm Fe}$ through $Z_{\rm Si}/Z_{\rm Fe}$ to be 1.4--1.7,
and $Z_{\rm S}/Z_{\rm Fe}$ through $Z_{\rm Ca}/Z_{\rm Fe}$ and
$Z_{\rm Ni}/Z_{\rm Fe}$ to all be $\sim$ 0.7--0.9.
Most of our observed abundance ratios are consistent (or at least
roughly consistent) with these predictions.
The current burst of star formation is estimated to have started at least 50 Myr ago 
(e.g., Dufour \et\ 1979).
Assuming star formation at a uniform rate of 30 $\Msun$ yr$^{-1}$ (Telesco 1978
measured a star formation rate 10 times that of the Milky Way's spiral arms),
this is sufficient 
to enrich the gas to the current metallicity levels within that time
span (M.\ Loewenstein, priv.\ comm.), although we cannot know for
certain how much enrichment of the gas has occurred within the last 10$^8$ yr 
and how much occurred closer in time to the merger.

Finally, we discuss the argument of Hardcastle, Evans and Croston (2007) that
"high-excitation" sources, which have torus-like circumnuclear absorbing gas 
and which accrete at moderate fractions of the Eddington limit (e.g., FR IIs), 
are fueled by a supply of cold gas, i.e., gas transported to the nucleus
as a result of a merger or tidal interactions. In contrast, "low-excitation" 
sources, which include all FR I's and some of the less powerful FR IIs, may be 
powered by accretion of hot (X-ray emitting) gas. As mentioned before, Cen A, 
although classified as an FR I, has heavy X-ray absorption characteristic of most 
FR IIs. Our observation of high metallicity, indicative of cold gas being 
transported inwards to the nucleus as a result of the merger, is thus consistent 
with this argument of "cold-mode" accretion.

\section{Conclusions}

The combination of {\it Suzaku}'s XIS and HXD has allowed us to obtain a high-quality spectrum of 
the nucleus (inner 2 kpc) of Cen A spanning 0.3--250 keV. The long exposure time, narrow response of the XIS,
and low background make the XIS spectrum one of the highest quality CCD spectra of Cen A to date, 
particularly below 2 keV. The HXD PIN above 12 keV is more sensitive than {\it RXTE} HEXTE or {\it INTEGRAL} 
instruments, allowing us to accurately and directly measure the 12--76 keV flux and spectral shape. The 
HXD GSO provides a direct estimate of the 45--250 keV flux of Cen A.

Our best-fit model includes several broadband components. The $>$3 keV emission can be fit by two absorbed power laws.
The primary (brighter) power law is absorbed by a column of about 1.5$\times$10$^{23}$ cm$^{2}$.
The secondary power law, with a brightness about 19$\%$ of the primary,
is absorbed by a much higher column, about 7$\times$10$^{23}$ cm$^{-2}$.
In our best-fit model, the photon indices were the same, and the best-fit value was 1.817$^{+0.023}_{-0.010}$.
Untying the photon indices did not result in a significant fit improvement.
Including data down to 0.3 keV, we find that two {\sc VAPEC}
components plus a third absorbed power-law component
can fit the data. The soft X-rays reveal emission lines due to
He- and H-like O, Fe {\sc XVII}, He- and H-like Ne, and  He-like Mg.
The strong Fe L emission supports an origin in collisionally ionized, 
rather than photoionized, gas. Two {\sc VAPEC} components of temperatures $k_{\rm B}T$ = 0.2 and 0.6 keV
can model the soft X-ray emission lines well.

The {\it Suzaku} data are consistent with scenarios previously forwarded to explain the nature of the 
various power-law components. As suggested by Evans \et\ (2004), 
the primary, hard X-ray power law is likely associated with Bondi accretion at the black hole.
The secondary power law could represent X-ray emission associated with the pc-scale VLBI jet,
although this is not certain. Alternatively, a partial-covering model, as suggested by Turner \et\ (1997), may apply:
84$\%$ (16$\%$) of the sky as seen from the X-ray source is obscured by a column 1.5 (7) $\times$10$^{23}$ cm$^{2}$.
In any event, it more likely that the material obscuring the hard X-rays is associated
with the circumnuclear material than with the optical dust lane.
The third, soft X-ray power-law component is consistent with absorption by
the optical dust lane only. It likely represents a blend of
diffuse emission and knot emission from the kpc-scale jet (as well as possibly the pc-scale jet),
plus emission from a population of XRBs and other X-ray point sources,
plus a likely contribution from scattered nuclear emission, which is likely less than 1$\%$
of the total hard X-ray power-law emission. 

The hard X-ray photon index, the 2--10 keV flux, the Fe K$\alpha$ line intensity
and the column density of the material
absorbing the primary power law are all consistent with recent
observations over the last $\sim$5 years,
suggesting that no large variations in these parameters have occurred over the last few years.
During the {\it Suzaku} observation, the hard X-ray flux
was not strongly variable on short timescales, showing only a 10$\%$ increase in flux
throughout the observation.

K-shell absorption edges due to Fe, Ca, and S are significantly detected.
Several fluorescent emission lines, likely originating in 
the absorbing material, 
are detected. In addition to Fe K$\alpha$ and Fe K$\beta$, 
K$\alpha$ lines from Si, S, Ar, Ca and Ni are detected at 
at least 90$\%$ confidence. The latter three are detected
for the first time in Cen A. The strict upper limit on the amount of 
Compton reflection, $R < 0.05$, supports the notion that
the fluorescent lines originate in Compton-thin material.
The Fe K$\alpha$ line width is $<$ 2500 km s$^{-1}$ FWHM.
If the line-emitting material is in virial orbit,
then it is no closer than 200 light-days to the black hole.
No emission due to He- or H-like Fe is detected. The Fe K edge energy
is consistent with absorption by Fe {\sc I}; ionization stages 
above Fe $\sim${\sc V} are ruled out. 
The gas does not likely completely surround the X-ray continuum source;
the equivalent width of the Fe K$\alpha$ line is more consistent
with gas which is clumpy and/or covers the continuum source
only partially, or the bulk of the gas may lie off the line of sight to the 
X-ray source, possibly in the form of a disk.

Metallicities are measured for the circumnuclear
absorbing gas for the first time in Cen A. 
The high signal-to-noise ratio of the XIS spectrum has
enabled us to use the K-shell absorption edge depths,
relative fluorescent line strengths, and
thermal plasma emission lines to derive
abundances (we have assumed that abundances in the 
thermal plasma gas are the same as those for the 
circumnuclear absorbing material).
We observe a value of 
[m/H] = log($Z_{\rm Fe}$) = +0.1 for the circumnuclear material/diffuse plasma
which is much higher than for the halo stars at 20--30 kpc radii.
It is likely that the merger that created Cen A's famous
edge-on dust disk also triggered the infall of cold gas into the inner regions of Cen A, where it 
currently accretes onto the black hole. 
Our metallicity observation suggests that the accreting material could not have originated
in the relatively metal-poor outer halo unless
enrichment due to local star formation occurred at some point.
The relative observed abundances are indeed consistent with enrichment
by a mixture of Type II and Type Ia supernovae.



This high signal-to-noise ratio CCD observation 
has yielded breakthroughs in constraining abundances,
but further progress can be made
with a very high spectral resolution instrument, such as the
planned calorimeter aboard {\it Constellation-X}.
By better constraining the 
widths and intensities of K$\alpha$ emission lines
and soft X-ray emission lines, as well as
edge energies and depths,
{\it Constellation-X} will thus yield
abundance constraints on the
circumnuclear material in potentially many dozens of AGNs once it is launched.


\acknowledgements The authors gratefully acknowledge the
dedication and hard work of the {\it Suzaku} hardware teams
and operations staff for making this observation possible and
for assistance with data calibration and 
analysis. The authors thank the referee for useful comments
and suggestions. A.M.\ thanks Rick Rothschild, Mike Loewenstein, and 
Tim Kallman for useful discussions. 
This research has made use of HEASARC online services, supported by
NASA/GSFC. This research has also made use of the NASA/IPAC 
Extragalactic Database,
operated by JPL/California Institute of Technology, under
contract with NASA.

\clearpage


\begin{deluxetable}{lccc}
\tabletypesize{\footnotesize}
\tablewidth{6.5in}
\tablenum{1}
\tablecaption{Soft X-ray Emission Lines\label{tab1}}
\tablehead{
\colhead{Line}    & \colhead{Observed Energy} & \colhead{Intensity}                                  & \colhead{$EW$}     \\
\colhead{Identification}& \colhead{Centroid (keV)}  & \colhead{(10$^{-5}$ ph cm$^{-2}$ s$^{-1}$)} & \colhead{(eV)}   }
\startdata
O {\sc VII} &  0.564$^{+0.006}_{-0.004}$ & 10.9 $\pm$ 2.1      &  39$\pm$0.8     \\
O {\sc VIII} & 0.654$\pm$0.003           & 9.3$^{+2.6}_{-1.1}$ &  37$^{+10}_{-4}$ \\
Fe {\sc XVII} 3s--2p ($^3$$P_1$/$^3$$P_2$/$^1$$P_1$)     & 0.732$^{+0.004}_{-0.006}$ & 5.3$^{+1.0}_{-0.7}$ & 23$^{+4}_{-3}$   \\
Fe {\sc XVII} 3d--2p ($^1$$P_1$/$^3$$D_1$) & 0.820$^{+0.005}_{-0.003}$ & 7.4$^{+0.8}_{-1.4}$ & 33$^{+4}_{-6}$   \\
Ne {\sc IX}   & 0.901$\pm$0.004           & 6.3$^{+0.6}_{-0.9}$ & 36$^{+3}_{-5}$  \\
Ne {\sc X}    & 1.010$\pm$0.003           & 4.1$\pm$0.6         & 33$\pm$5  \\
Mg {\sc XI}   & 1.331$^{+0.008}_{-0.004}$ & 1.5$\pm$0.4         & 22$\pm$6   \\
\enddata
\tablecomments{Equivalent widths $EW$ (col.\ [4]) were determined relative to a locally-fit continuum.}
\end{deluxetable}


\begin{deluxetable}{lcccc}
\tabletypesize{\footnotesize}
\tablewidth{6.5in}
\tablenum{2}
\tablecaption{Best-fit parameters for selected models \label{tab2}}
\tablehead{
\colhead{Parameter} & \colhead{Model 5}        & \colhead{Model 6} & \colhead{Model 7} & \colhead{Model 8}}
\startdata
$\chi^2$/$dof$                    &   9806.8/9158                      & 9745.1/9159                 &  9639.6/9144              & 9529.9/9142 \\
$\Gamma_{1,2}$                    & 1.852$^{+0.008}_{-0.019}$          & 1.831$^{+0.019}_{-0.010}$   & 1.820$^{+0.016}_{-0.010}$ & 1.817$^{+0.023}_{-0.010}$ \\
PL1 normalization$^1$                     & 0.116$^{+0.002}_{-0.012}$          & 0.112$\pm$0.002             & 0.111$\pm$0.002  & 0.111$^{+0.001}_{-0.002}$  \\
$N_{\rm H,1}$ (10$^{22}$ cm$^2$) &    15.2$^{+0.1}_{-0.6}$            &    15.1$\pm$0.2             &    14.7$\pm$0.2  & 14.7$^{+0.3}_{-0.2}$   \\
PL2 normalization$^1$                     &  0.026$^{+0.004}_{-0.002}$         & 0.022$\pm$0.003             & 0.021$^{+0.002}_{-0.003}$ & 0.021$\pm$0.003   \\
$N_{\rm H,2}$ (10$^{22}$ cm$^2$) &   82$\pm$7                         & 75$\pm$8                    &      72$^{+10}_{-3}$    & 70$^{+11}_{-7}$  \\
{\sc vapec}1 $k_{\rm B}T$ (keV)    &   0.31$^{+0.03}_{-0.02}$        &   0.24$\pm$0.02             &   0.23$\pm$0.02   & 0.22$^{+0.02}_{-0.01}$ \\
{\sc vapec}1 normalization                 & 7.2$^{+0.1}_{-1.3}\times$10$^{-4}$ & 5.0$^{+0.1}_{-0.3}\times$10$^{-4}$  & 5.0$^{+0.1}_{-0.3}\times$10$^{-4}$ &  4.3$^{+0.1}_{-0.3}$$\times$10$^{-4}$ \\
{\sc vapec}2 $k_{\rm B}T$ (keV)    &     0.74$^{+0.09}_{-0.03}$         & 0.62$\pm$0.01  & 0.62$\pm$0.01  & 0.56$^{+0.01}_{-0.03}$  \\
{\sc vapec}2 normalization                 & 3.7$^{+0.3}_{-0.8}\times$10$^{-4}$ & 5.1$\pm$0.3$\times$10$^{-4}$ & 5.2$^{+0.2}_{-0.3}\times$10$^{-4}$ & 4.8$\pm$0.4$\times$10$^{-4}$ \\
$\Gamma_3$                        & 1.16$^{+0.13}_{-0.10}$             & 1.28$^{+0.08}_{-0.12}$    &  1.41$^{+0.14}_{-0.09}$  &  1.31$^{+0.08}_{-0.10}$ \\
PL3 normalization$^1$                     & 6.9$^{+0.5}_{-0.3}\times10^{-4}$   & 7.4$^{+0.4}_{-0.5}\times10^{-4}$ & 7.8$^{+0.5}_{-0.9}\times10^{-4}$ & 7.0$^{+0.6}_{-0.3}\times10^{-4}$  \\
$N_{\rm H,3}$ (10$^{22}$ cm$^2$) & 0(fixed)                           &  $<$0.03               & $<$0.05    & $<$0.03 \\   
\enddata
\tablecomments{Best-fit parameters for selected broadband models. All models include two absorbed X-ray power laws (PL1, PL2),
two {\sc vapec} components, a third soft X-ray power law (PL3), and Fe K$\alpha$ and K$\beta$ emission lines.
Listed are:
Model 5, where PL3 is not absorbed;
Model 6, where PL3 is absorbed;
Model 7, where {\sc zvphabs} abundances are thawed for S, Ca, and Fe and fluorescent lines for 
Si, S, Ar, Ca, and Ni are added; and
Model 8, our best-fit model, in which the {\sc vapec} abundances $Z_{\rm Ne}$ = 
2.7$\pm$0.3 and $Z_{\rm Mg}$ = 1.9$\pm$0.4.  \\
$^1$: units are ph keV$^{-1}$ cm$^{-2}$ s$^{-1}$ at 1 keV. }
\end{deluxetable}


\begin{deluxetable}{lcccccc}
\tabletypesize{\footnotesize}
\tablewidth{6.5in}
\tablenum{3}
\tablecaption{Fluorescent Emission Line Parameters\label{tab3}}
\tablehead{
\colhead{}     & \colhead{Energy}         & \colhead{Intensity (10$^{-5}$ }  & \colhead{$EW_{\rm obs}$} & \colhead{} & \colhead{$F$-test} & \colhead{$EW_{\rm calc}$} \\
\colhead{Line} & \colhead{Centroid (keV)} & \colhead{ph cm$^{-2}$ s$^{-1}$)} & \colhead{(eV)}   & \colhead{$\Delta\chi^2$} & \colhead{Prob.}   & \colhead{(eV)} }
\startdata
Si K$\alpha$           & 1.71$\pm$0.01   &  0.9$\pm$0.2 & 24$\pm$6 & 39.8 & 7.6$\times$10$^{-10}$   & 26.3 \\
S K$\alpha$            & 2.307$\pm$0.016 &  0.8$\pm$0.4 &  6$\pm$2 &  5.0 & 9.6$\times$10$^{-2}$    & 11.4 \\
Ar K$\alpha$           & 2.994$\pm$0.023 &  1.2$\pm$0.5 & 13$\pm$4 & 20.1 & 8.0$\times$10$^{-5}$    & 4.7  \\
Ca K$\alpha$           & 3.690$\pm$0.023 &  1.8$\pm$0.7 &  8$\pm$3 &  6.2 & 5.4$\times$10$^{-2}$    & 3.7  \\
Fe K$\alpha$           & 6.394$\pm$0.003 & 23.2$\pm$1.0 & 83$\pm$3 & 3077 & $<$1$\times$10$^{-40}$  & 128  \\
Fe K$\beta$            & 7.056$^{1}$     &  3.0$^2$     & 11$^2$   & 28.3 & 2.7$\times$10$^{-7}$ &  17$^2$\\
Ni K$\alpha$           & 7.47$\pm$0.05   &  1.4$\pm$0.8 & 15$\pm$5 &  9.3 & 1.1$\times$10$^{-2}$     & 7.1   \\
\enddata
\tablecomments{Results are for Model 8.
Observed equivalent widths (col.\ [4]) are determined relative to a locally-fit continuum.
All widths were tied to that for the Fe K$\alpha$ line, 14$^{+10}_{-14}$ eV.
$EW_{\rm calc}$ (col.\ [7]) denotes predicted equivalent width values using equation 1 (uniform covering
assuming line of sight column density), solar abundances, and a covering fraction of 1.0. \\
$^1$ denotes a fixed parameter. \\
$^2$ denotes value tied to 0.13 that for the Fe K$\alpha$ line.}
\end{deluxetable}



\begin{deluxetable}{lccc}
\tabletypesize{\footnotesize}
\tablewidth{6.5in}
\tablenum{4}
\tablecaption{Hard X-ray Absorption Edges\label{tab4}}
\tablehead{
\colhead{K-shell} & \colhead{{\sc zvphabs}} & \colhead{Edge Energy} &  \colhead{Edge Optical} \\
\colhead{Edge} & \colhead{Abundance}     & \colhead{(keV)}       &  \colhead{Depth, $\tau$}}
\startdata
S                 &   1.14$^{+0.11}_{-0.15}$ & 2.468$^{+0.014}_{-0.011}$ & 0.24$^{+0.03}_{-0.06}$  \\
Ca                &   1.5$^{+0.8}_{-0.6}$    & 4.040$\pm$0.045           & 0.03 $\pm$ 0.01         \\
Fe                &   1.17$^{+0.12}_{-0.09}$ & 7.103$\pm$0.015           & 0.21 $\pm$ 0.01         \\
\enddata
\tablecomments{The {\sc zvphabs} abundances in col.\ (2) are from Model 8. The edge energy and optical depths
(cols.\ [3]--[4]) are from a model with {\sc zvphabs} abundances for Fe, Ca and S set to zero and edges
fit instead.}
\end{deluxetable}   


\begin{deluxetable}{lcc}
\tabletypesize{\footnotesize}
\tablewidth{6.5in}
\tablenum{5}
\tablecaption{Abundance Ratios\label{tab5}}
\tablehead{ \colhead{Method} &  \colhead{Ratio} &  \colhead{Value}} 
\startdata
1 & $Z_{\rm O}/Z_{\rm Fe}$   &   0.8 $\pm$ 0.4 \\
1 & $Z_{\rm Ne}/Z_{\rm Fe}$  &   2.3 $\pm$ 0.3 \\           
1 & $Z_{\rm Mg}/Z_{\rm Fe}$  &   1.6 $\pm$ 0.3 \\              
3 & $Z_{\rm Si}/Z_{\rm Fe}$  &     1.5 $\pm$ 0.4  \\
2 & $Z_{\rm S}/Z_{\rm Fe}$   &   1.0 $\pm$ 0.1 \\ 
3 & $Z_{\rm S}/Z_{\rm Fe}$   &     0.8 $\pm$ 0.3 \\
3 & $Z_{\rm Ar}/Z_{\rm Fe}$  &       4 $\pm$ 1 \\
2 & $Z_{\rm Ca}/Z_{\rm Fe}$  &   1.3 $\pm$ 0.6 \\
3 & $Z_{\rm Ca}/Z_{\rm Fe}$  &       3 $\pm$ 1 \\
3 & $Z_{\rm Ni}/Z_{\rm Fe}$  &       3 $\pm$ 1 \\
\enddata
\tablecomments{Methods 1, 2, and 3 (col.\ [1]) denote abundances derived from
thermal plasma emission lines, K shell edge depths, and K$\alpha$ fluorescent
emission line intensities, respectively.}
\end{deluxetable}

\clearpage


\begin{figure}
\epsscale{0.60}
\plotone{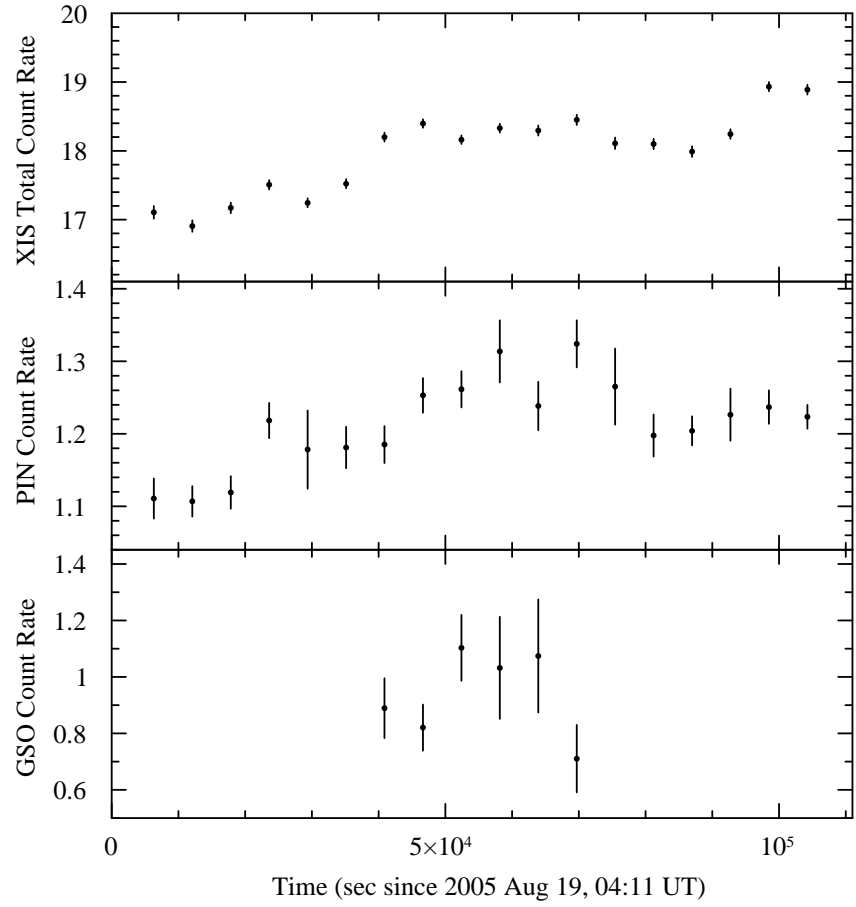}
\caption{Orbitally binned XIS and HXD light curves. The top panel shows the 
2--10 keV count rate light curve, summed over all four XIS cameras.
The 12--76 keV HXD PIN light curve is shown in the middle panel.
The 45--250 keV HXD GSO light curve is shown in the bottom panel.}
\end{figure}



\begin{figure}
\epsscale{0.80}
\plotone{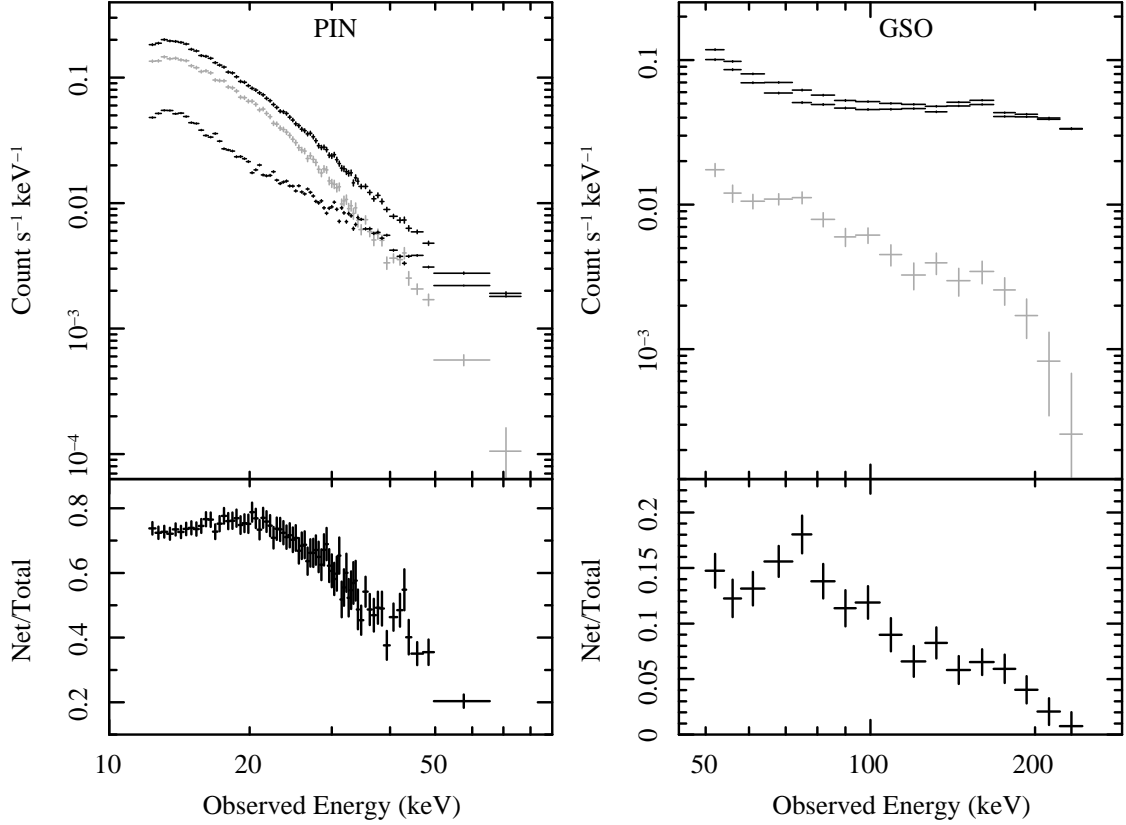}   
\caption{HXD PIN ({\it left}) and GSO ({\it right}) spectra. The upper panels show
the net source spectrum ({\it gray points}), the background ({\it lower black points}),
and the total (source + background) spectrum ({\it upper black points}).
The PIN spectra have been binned such that the net spectrum has a minimum
signal-to-noise ratio of 10$\sigma$ per bin.
The GSO spectra have been binned as described in $\S$2.
The lower panels shows the ratio of the net source spectrum to
the total spectrum.}
\end{figure}



\begin{figure}
\epsscale{0.80}
\plotone{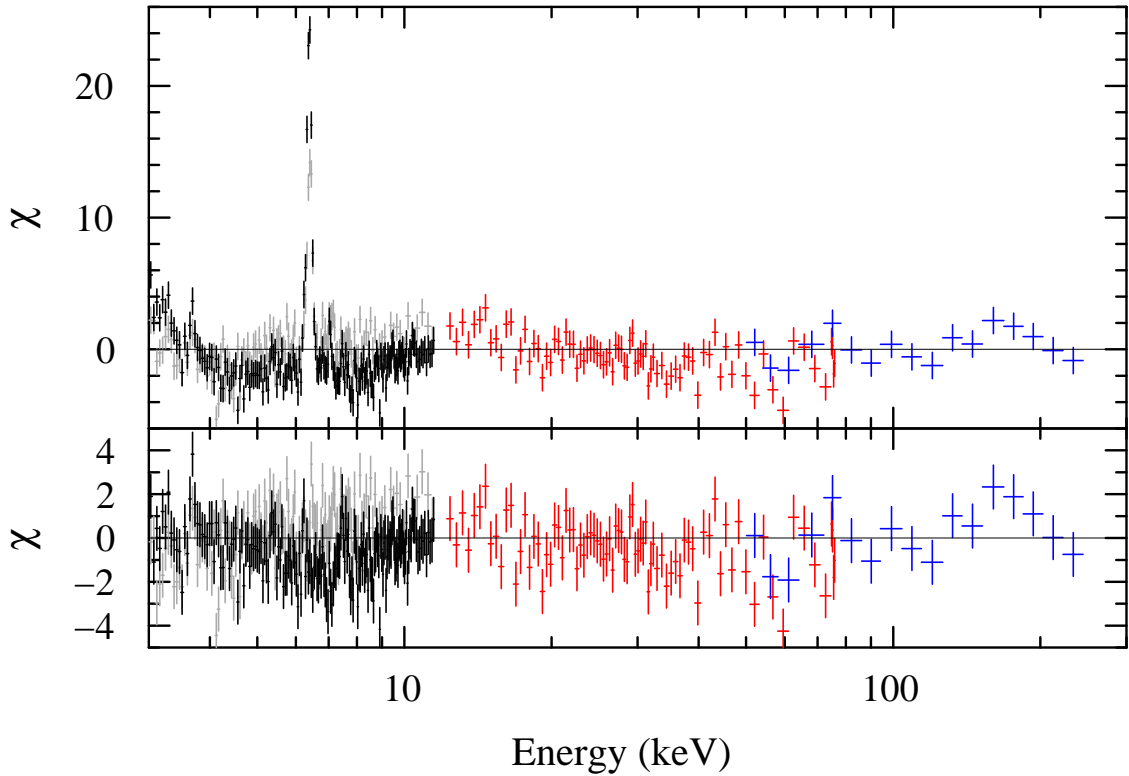} 
\caption{Residuals to Model 1, the single absorbed power law ({\it top},
and Model 2, which included dual absorbed power-law components and 
Fe K$\alpha$ and K$\beta$ line emission ({\it bottom}). Black points denote
the three FI XIS spectra, which have been co-added here for clarity.
Gray, red and blue points denote XIS1, PIN, and GSO,
respectively. Rest-frame energies are shown.
The XIS data have been plotted with a binning factor of 10.}
\end{figure}



\begin{figure}
\epsscale{0.80}
\plotone{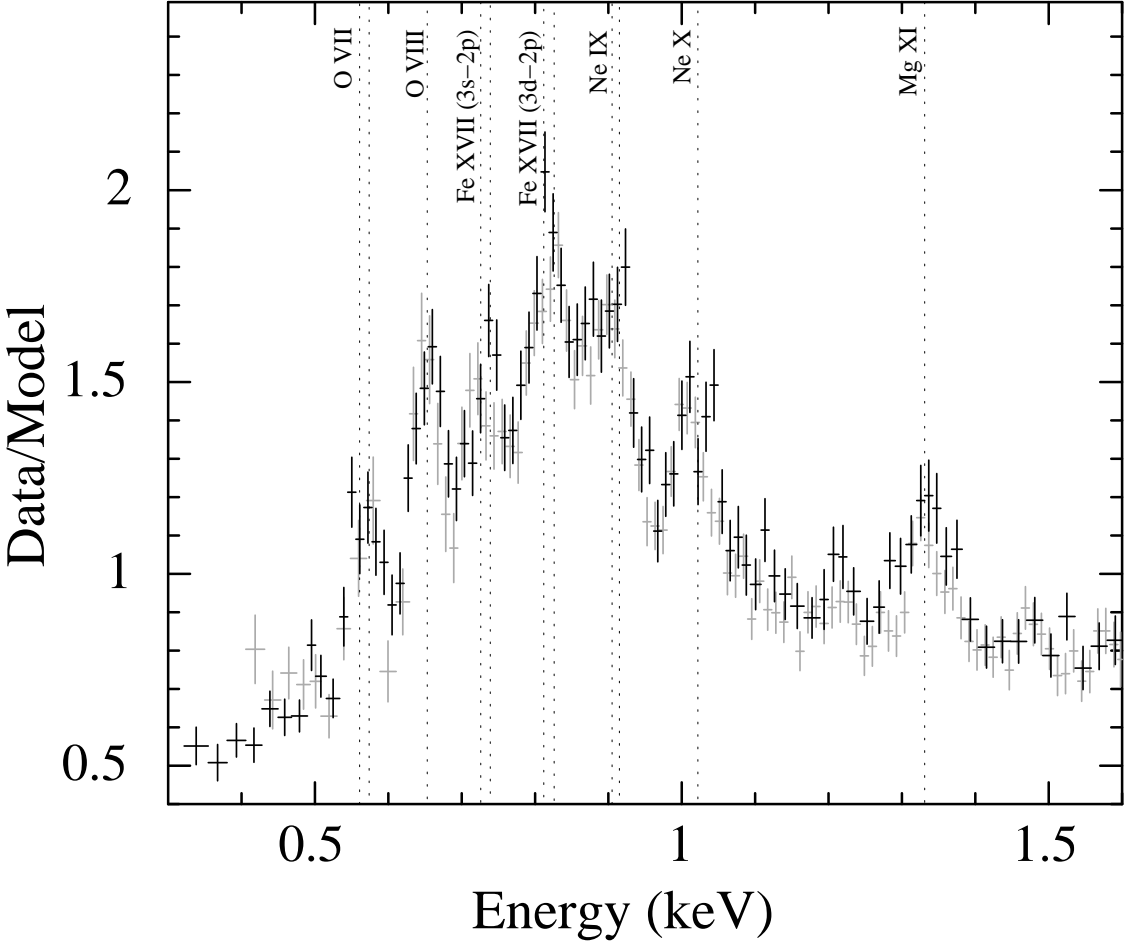}    
\caption{Ratio of the soft band data to a simple power law, 
showing the prominent emission lines.
Black points denote the XIS BI; gray points denote data from XIS0, 2 and 3,
which have been co-added for clarity.
Rest-frame energies are shown.
The data have been plotted with a binning factor of 3.}
\end{figure}



\begin{figure}
\epsscale{0.80}
\plotone{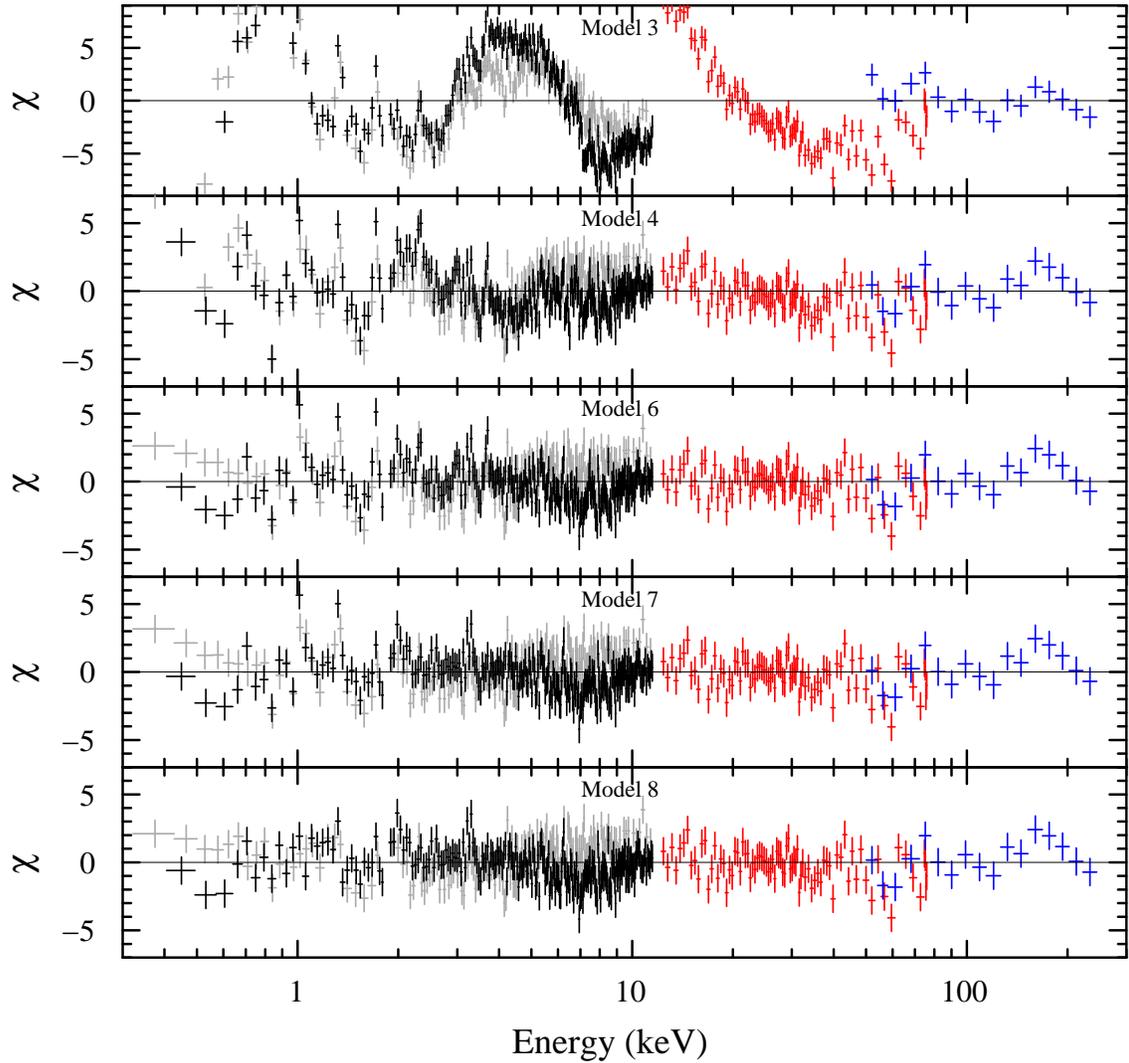}   
\caption{Residuals to various broadband models are shown. 
From top to bottom are: Model 3, with soft X-rays fit with a simple power law;
Model 4, with soft X-rays fit with two {\sc vapec} components only;
Model 6, with soft X-rays fit with two {\sc vapec} components plus an absorbed power law;
Model 7, with K-shell absorption edges and K$\alpha$ emission lines for
S, Si, Ar, Ca and Ni fit; and
Model 8, our best-fit model, with {\sc VAPEC} abundances for Ne and Mg thawed. 
Black points denote the three FI XIS spectra, which have been co-added here for clarity.
Gray, red and blue points denote XIS1, PIN, and GSO,
respectively. Rest-frame energies are shown.
The XIS data have been plotted with a binning factor of 10.}
\end{figure}



\begin{figure}
\epsscale{0.80}
\plotone{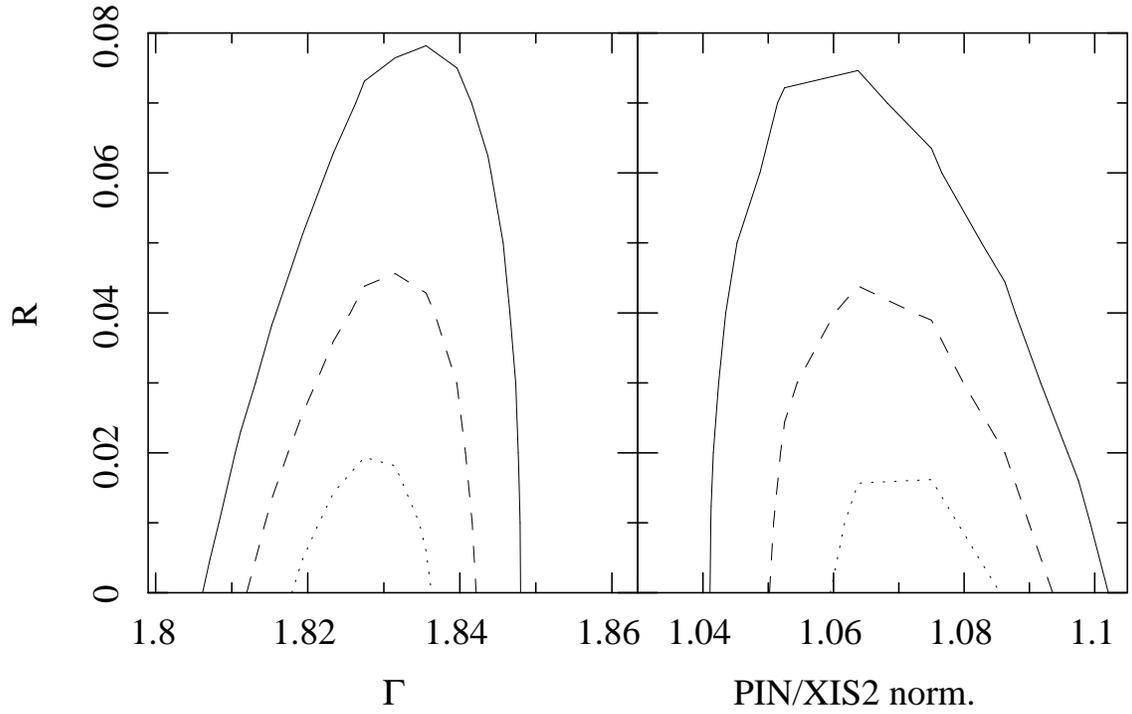}   
\caption{Contour plots of the Compton reflection fraction $R$
versus the hard X-ray photon index $\Gamma_1$=$\Gamma_2$
and the PIN/XIS2 instrument normalization for Model 8.
Dotted, dashed, and solid lines denote 68, 95.4, and 99.73$\%$
confidence levels, respectively.}
\end{figure}



\begin{figure}
\epsscale{0.60}
\plotone{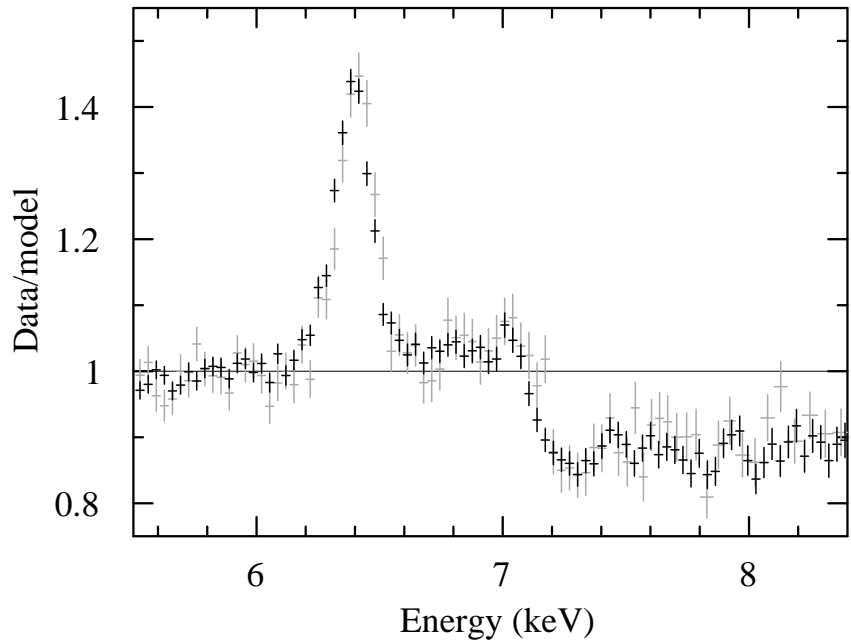}    
\caption{Residuals to a simple power-law model in the
Fe K bandpass, showing the prominent Fe K$\alpha$
emission line as well as the
Fe K$\beta$ line at 7.08$\pm$0.03 keV, the
Ni K$\alpha$ line at 7.47$\pm$0.05 keV, and the Fe K edge.
Black points denote the three FI XIS spectra, which have been co-added
for clarity; gray points denote XIS1.
Rest-frame energies are shown.
All data have been plotted with a binning factor of 10.}
\end{figure}



\begin{figure}
\epsscale{0.60}
\plotone{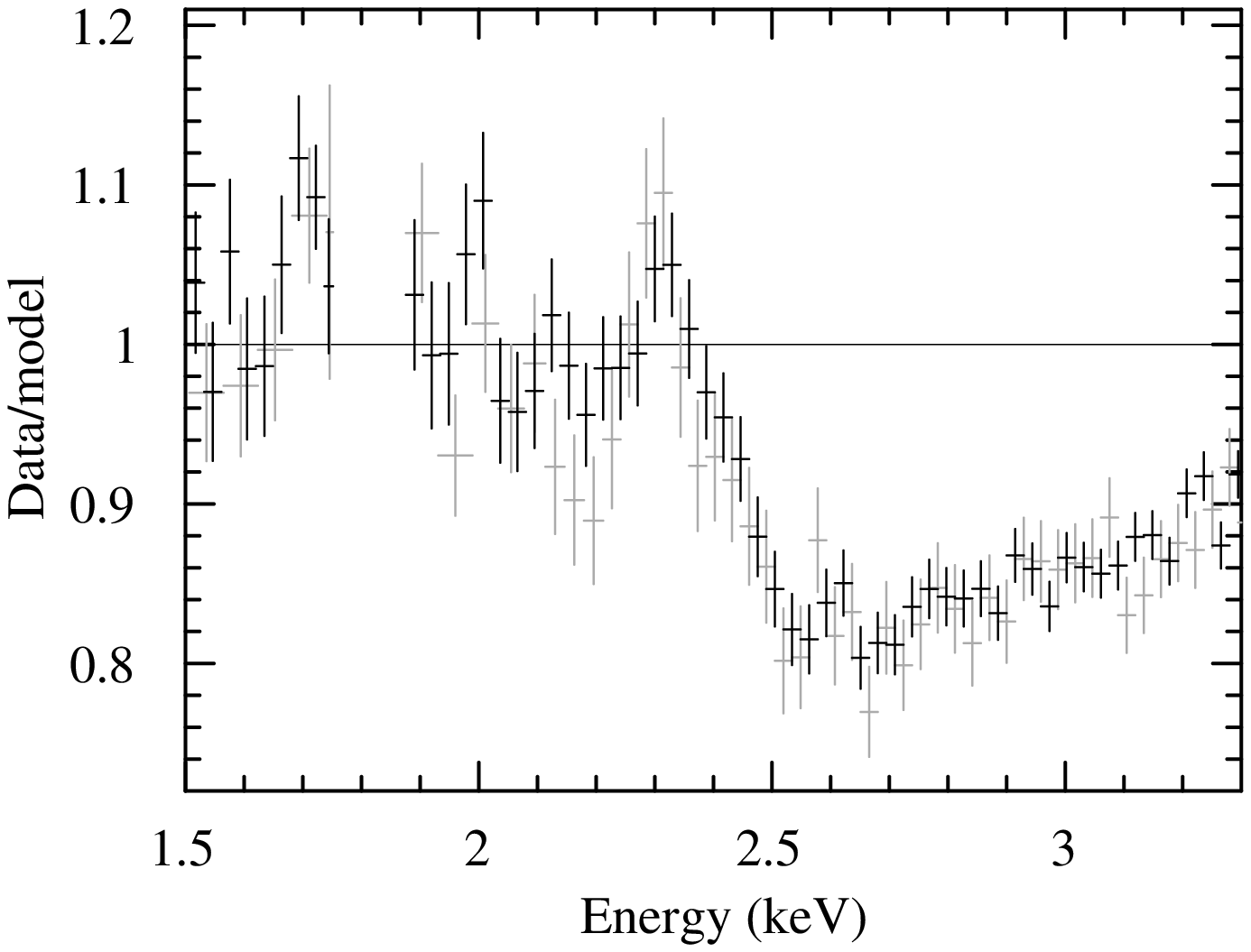}    
\caption{Residuals to a simple power-law model in the
S K and Si K bandpass, showing the Si K$\alpha$
emission line at 1.71$\pm$0.01 keV, the S K$\alpha$
line at 2.307$\pm$0.016 keV, and the S K edge.
Black points denote the three FI XIS spectra, which have been co-added
for clarity; gray points denote XIS1.
Rest-frame energies are shown.
All data have been plotted with a binning factor of 10.}
\end{figure}



\begin{figure}
\epsscale{0.60}
\plotone{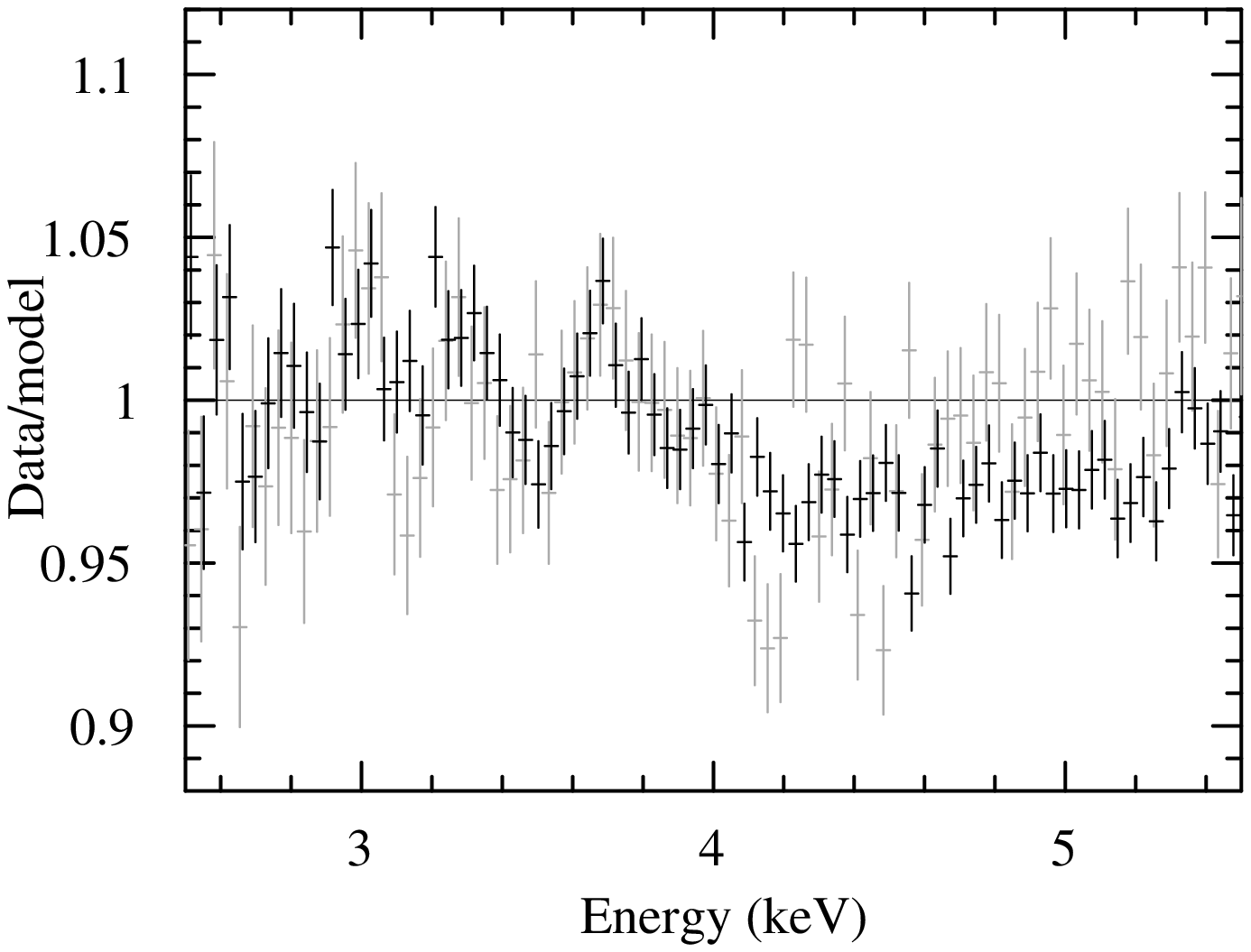}    
\caption{Residuals to a simple power-law model in the
Ar K and Ca K bandpass, showing the Ar K$\alpha$
emission line at 2.994$\pm$0.023 keV, the Ca K$\alpha$
line at 3.690$\pm$0.023 keV, and the Ca K edge.
Black points denote the three FI XIS spectra, which have been co-added
for clarity; gray points denote XIS1.
Rest-frame energies are shown.
All data have been plotted with a binning factor of 10.}
\end{figure}



\begin{figure}
\epsscale{0.80}
\plotone{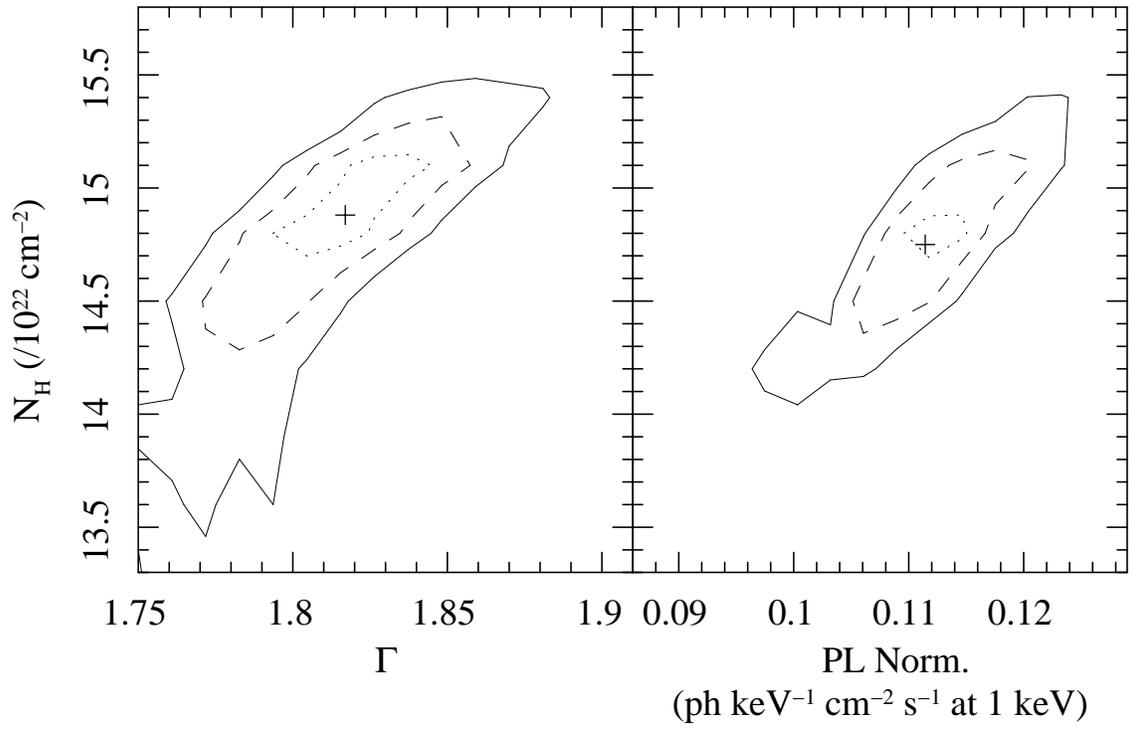}    
\caption{Contour plots of the line of sight column density absorbing
the primary power law versus the photon index ({\it left}) and
1 keV normalization ({\it right}) 
of the primary power-law component in Model 8. 
Dotted, dashed, and solid lines denote 68, 95.4, and 99.73$\%$
confidence levels, respectively. The plus signs mark the best-fit
values.}
\end{figure}



\begin{figure}
\epsscale{0.60}
\plotone{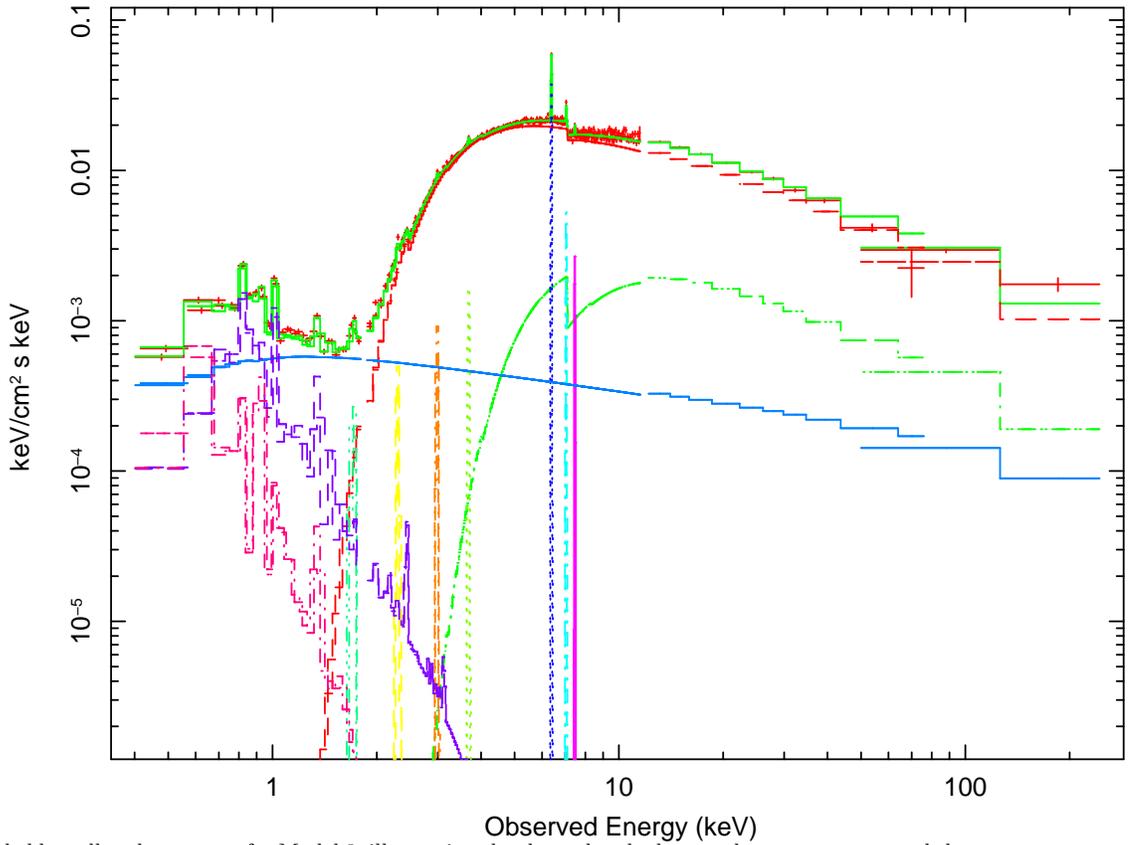}    
\caption{Unfolded broadband spectrum for Model 8, illustrating the 
three absorbed power-law components and the two {\sc vapec}
components. All data have been plotted with a binning factor of 10.}
\end{figure}



\begin{figure}
\epsscale{0.80}
\plotone{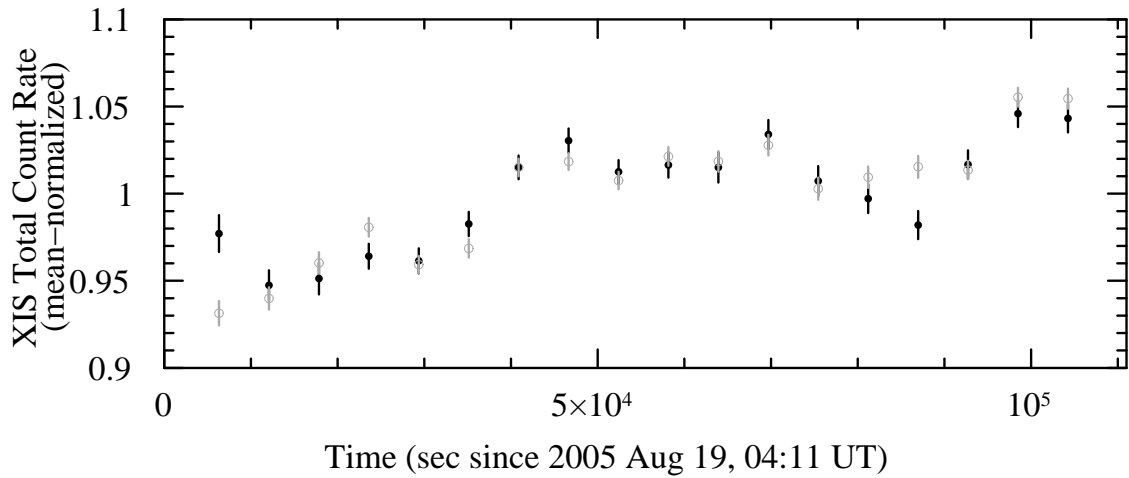}    
\caption{Mean-normalized, orbitally-binned 
2--4 keV ({\it black filled circles}) and 5--10 keV ({\it gray open circles})
light curves, summed over all four XISes.
Both light curves display very similar variability trends.}
\end{figure}

\end{document}